\documentclass[aps,prb,twocolumn,natbib,showpacs,floatfix,superscriptaddress, longbibliography]{revtex4-2}

\usepackage{graphicx}
\usepackage{amsmath,amsthm,amssymb,amsfonts,bbold,bm,color,mathtools}
\usepackage{float}
\usepackage{epstopdf}
\usepackage{hyperref}
\usepackage{enumerate}
\usepackage{wasysym}
\usepackage{url}
\usepackage{dsfont}
\usepackage{braket}
\usepackage{comment} 
\usepackage[capitalize]{cleveref}
\usepackage[normalem]{ulem}
\usepackage{xcolor}
\hypersetup{colorlinks=true,linkcolor=blue,citecolor=blue,urlcolor=blue}

\newcommand{\cT}{{\cal T}}

\newcommand{\cP}{{\cal P}}

\newcommand{\cE}{{\cal E}}
\newcommand{\cF}{{\cal F}}

\newcommand{\pf}{{45^{\circ}}}

\definecolor{greenish}{rgb}{0,0.8,0.4}

\begin{document}

\title{Josephson diode effects in twisted nodal superconductors}

\author{Pavel A. Volkov}
\affiliation{Department of Physics, Harvard University, Cambridge, Massachusetts, 02138 USA}
\affiliation{Department of Physics, University of Connecticut, Storrs, Connecticut 06269, USA}
\affiliation{Department of Physics and Astronomy, Center for Materials Theory, Rutgers University, Piscataway, NJ 08854, USA}

\author{\'Etienne Lantagne-Hurtubise}
\affiliation{Department of Physics and Institute for Quantum Information and Matter,
California Institute of Technology, Pasadena, California 91125, USA}

\author{Tarun Tummuru}
\affiliation{Department of Physics and Astronomy \& Stewart Blusson Quantum Matter Institute, University of British Columbia, Vancouver, BC V6T 1Z4, Canada}
\affiliation{Department of Physics, University of Zurich, Winterthurerstrasse 190, Zurich 8057, Switzerland}

\author{Stephan Plugge}
\affiliation{Instituut-Lorentz, Universiteit Leiden, P.O. Box 9506, 2300 RA Leiden, The Netherlands}

\author{J. H. Pixley}
\affiliation{Department of Physics and Astronomy, Center for Materials Theory, Rutgers University, Piscataway, NJ 08854, USA}
\affiliation{
Center for Computational Quantum Physics, Flatiron Institute, 162 5th Avenue, New York, NY 10010
}%

\author{Marcel Franz}
\affiliation{Department of Physics and Astronomy \& Stewart Blusson Quantum Matter Institute, University of British Columbia, Vancouver, BC V6T 1Z4, Canada}

\date{\today}

\begin{abstract}
   Recent Josephson tunneling experiments on twisted flakes of high-$T_c$ cuprate superconductor  Bi$_2$Sr$_2$CaCu$_2$O$_{8+x}$ revealed a non-reciprocal behavior of the critical interlayer Josephson current -- i.e., a Josephson diode effect. Motivated by these findings we study theoretically the emergence of the Josephson diode effect in twisted interfaces between nodal superconductors, and highlight a strong dependence on the twist angle $\theta$ and damping of the junction. In all cases, the theory predicts diode efficiency that vanishes exactly at $\theta = 45^\circ$ and has a strong peak at a twist angle close to $\theta = 45^\circ$, consistent with experimental observations. Near $45^\circ$, the junction breaks time-reversal symmetry $\cT$ spontaneously. We find that for underdamped junctions showing hysteretic behavior, this results in a \emph{dynamical} Josephson diode effect in a part of the $\cT$-broken phase. The direction of the diode is trainable in this case by sweeping the external current bias. This effect provides a sensitive probe of spontaneous $\cT$-breaking. We then show that explicit $\cT$-breaking perturbations with the symmetry of a magnetic field perpendicular to the junction plane lead to a {\em thermodynamic} diode effect that survives even in the overdamped limit. We discuss an experimental protocol to probe the double-well structure in the Josephson free energy that underlies the tendency towards spontaneous $\cT$-breaking even if $\cT$ is broken explicitly. Finally, we show that in-plane magnetic fields can control the diode effect in the short junction limit, and predict the signatures of explicit $\cT$-breaking in Shapiro steps.
\end{abstract}

\maketitle

\section{Introduction}

Twisted nodal superconductors have emerged as a promising platform to engineer exotic forms of superconductivity~\cite{Yang2018, Can2021, Volkov2020, Volkov2022, Song2022, Lu2022, Tummuru2022, Haenel2022}, capable of hosting topological phases with potentially large energy scales inherited from constituent high-$T_c$ materials such as monolayer cuprate  Bi$_2$Sr$_2$CaCu$_2$O$_{8+x}$ (BSCCO)~\cite{Yu2019}. In particular, time-reversal symmetry breaking (TRSB), a prerequisite for chiral topological superconductivity, has been predicted to occur spontaneously \cite{Can2021,Volkov2021} for twist angles $\theta$ close to $45^\circ$. 
The mechanism driving this unconventional transition is the second harmonic of the interlayer Josephson current-phase relation (CPR), that favors a non-trivial (i.e., different from 0 or $\pi$) phase difference across the junction in equilibrium \cite{Can2021,Volkov2021,Tummuru2021}.

Recent Josephson experiments \cite{Zhao2021} at the twisted interface between thin flakes of BSCCO have provided evidence for this second harmonic. In particular, anomalous Fraunhofer patterns in the presence of in-plane magnetic fields and fractional Shapiro steps under microwave driving are consistent with the second harmonic being dominant near $\theta = 45^{\circ}$.
However, neither of these observations are directly sensitive to TRSB at the interface. Moreover, it is possible theoretically to have dominant second harmonic in CPR that does not lead to spontaneous TRSB \cite{Can2021,Volkov2021,Tummuru2021}. Additional complication arises due to the two-dimensional nature of the interface where TRSB occurs, precluding bulk probes, such as muon spin resonance or nuclear spin resonance. Therefore, more direct probes of TRSB are required.

Some observables proposed to detect TRSB at the interfaces include polar Kerr effect measurements~\cite{Can2021b} and spontaneous currents around impurities or edges \cite{sigristueda,Kallin_2016}, that may be detected using SQUID magnetometry \cite{squid_review} or oscillations in magnetic field \cite{Wang2020}. Here we argue that a particularly sensitive probe of TRSB in superconductors is the Josephson diode effect (JDE), whereby the critical current in a Josephson junction becomes dependent on the current polarity.  Conceptually, this follows from the simple observation that current is odd under both time reversal $\cT$ and inversion $\cP$ and hence the diode effect can be present only when both symmetries are broken~\cite{Zinkl2021}.

Non-reciprocal transport properties of superconductors have been explored recently in a variety of setups where $\cP$ and $\cT$ are broken. In non-centrosymmetric superconductors, supercurrent diode effects can be induced by applying an external magnetic field or by proximitizing with a magnetic material. This was observed recently in an experiment on a Nb/V/Ta superlattice without an inversion center~\cite{Ando2020}, in magnetic Josephson junctions built from twisted bilayer graphene~\cite{diez2023symmetry,hu2023} or d-wave superconductors on top of topological insulators \cite{tanaka2022}, in InAs quantum wells~\cite{Baumgartner2021}, in NbSe$_2$ nanowires~\cite{Bauriedl2021} or films~\cite{Shin2021}, and in the Dirac semimetal NiTe$_2$~\cite{Pal2021}. Recent theoretical works on the JDE in inversion-broken superconductors under an external magnetic field, including Rashba superconductors, were also reported~\cite{He2021, Yuan2021, Daido2022, Zhang2021, Davydova2022}. 

SC diode effects can also occur through spontaneous $\cT$ breaking, which gives rise to a hysteresis loop in the diode response as a function of applied magnetic field, as reported in experiments on alternating-twist trilayer graphene~\cite{Lin2021zerofield, Scammell2022}. Another recent experiment on TMD-based junctions reports JDE without a clear source of TRSB, which is difficult to reconcile with the basic symmetry requirements mentioned above~\cite{Wu2022} and could be related to theory ideas developed in \cite{Chen2018}. Other relevant theory work on JDE include Refs.~\cite{Souto2022, Steiner2022, kochan2023phenomenological}.

\begin{figure}[t]
		\includegraphics[width=0.8\linewidth]{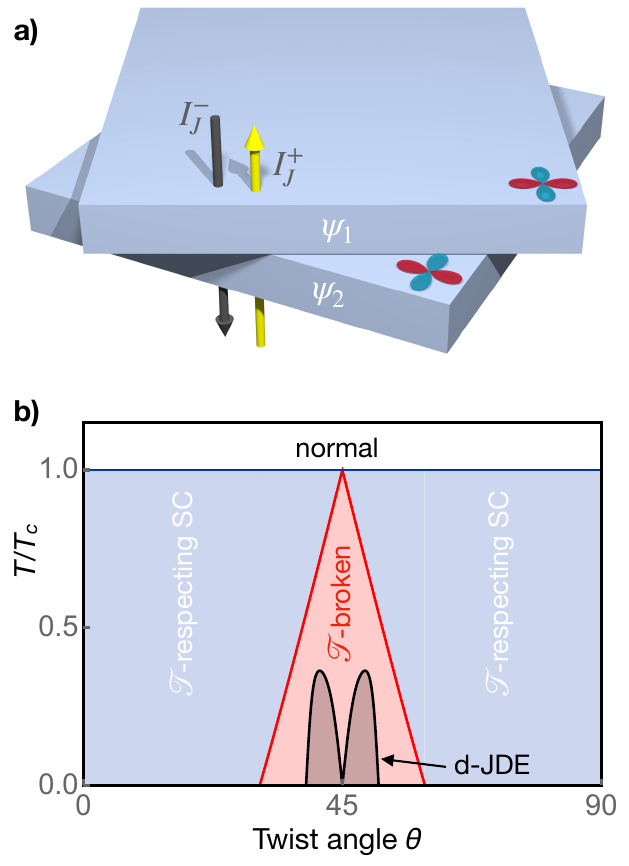}
	\caption{(a) Two flakes of a $d$-wave superconductor form a twisted $c$-axis Josephson junction. In the absence of time reversal $\cT$ and inversion $\cP$, the critical Josephson currents for the two current directions are allowed to be unequal. (b) Schematic phase diagram implied by the GL theory. Dynamical Josephson diode effect (d-JDE) is found to occur inside the spontaneously $\cT$-broken phase.
	}
	\label{fig:setup}
\end{figure}

In this work, we investigate the JDE in twisted $c$-axis cuprate junctions, schematically depicted in Fig.\ \ref{fig:setup}(a), with the aim of providing a theoretical background for recent experimental results~\cite{Zhao2023,Ghosh2022,Zhu2023}. In particular, Ref.\,\cite{Ghosh2022} has reported a diode effect induced and controlled by magnetic field, whereas Ref.~\cite{Zhu2023} reported a diode effect in the absence of magnetic field. Ref.\ \cite{Zhao2023}, in addition to the above effects, observed a zero-field diode effect that can be trained by a directed current sweeping.

We base our analysis on the mean-field theory of twisted $d$-wave superconductors following Refs.\ \cite{Can2021,Tummuru2021, Volkov2021}. This approach famously predicts a strong suppression of the Josephson critical current for junctions with a twist angle $\theta$ close to $45^\circ$. Importantly, for these values of the twist angle the remnant value of the critical current is predicted to be due to a second-harmonic Josephson effect generated by Cooper pair co-tunneling. Both of these features were for the first time clearly observed in recent experiment \cite{Zhao2021}. Data presented in Refs.~\cite{Zhu2021, Zhu2023} on nominally similar samples showed only moderate suppression and were interpreted as evidence for an $s$-wave pairing component. However, experiments that could distinguish the second Josephson harmonic close to $\theta=45^\circ$ have not been performed in those works, leaving open a possibility that the same physics is being realized.

Our analysis identifies two different pathways that can engender non-reciprocal currents in twisted $d$-wave junctions: (i) \emph{Dynamical} JDE brought about by the spontaneous TRSB and (ii) \emph{Thermodynamic} JDE relying on explicit TRSB unrelated to the Josephson physics. In the former the polarity of the diode depends on its history, is trainable with current biasing and allows for detection of spontaneous TRSB in Josephson experiments. The latter implies a `memory effect', whereby the junction exhibits the same polarity of JDE after is has been cycled above the superconducting critical temperature $T_c$ or driven into the resistive state by exceeding its critical current. 

Dynamical JDE arises as a result of the characteristic double-minimum structure of the free energy in the $\cT$-broken phase, predicted to occur in bilayer $d$-wave junctions in the vicinity of $\theta=45^\circ$. Importantly, because of kinematic and symmetry constraints the effect vanishes at $45^\circ$ twist and extends over a subregion of the $\cT$-broken phase as illustrated in the phase diagram Fig.\ \ref{fig:setup}(b). In the presence of explicit TRSB, on the other hand, the thermodynamic JDE occurs at all twist angles  except $\theta=0^\circ,45^\circ$ and at all temperatures below $T_c$. Its strength, as measured by the figure of merit $\eta$ defined below, is found to peak in the vicinity of $\theta=45^\circ$.

The rest of this paper is organized as follows. In Sec.~\ref{sec:Setup} we introduce the Ginzburg-Landau (GL) theory formalism appropriate for twisted nodal superconductors and discuss the relevant symmetries of the free energy in the context of the JDE. In Sec.~\ref{sec:dyndiode} we investigate the dynamical JDE and explain how it can be used as a sensitive probe of spontaneous $\cT$-breaking. In Sec.~\ref{sec:trsb} we consider the explicit breaking of $\cT$ and its consequences for both thermodynamic and dynamical diode effects. Detailed experimental protocols that allow to distinguish the two categories of JDEs are contrasted in Sec.~\ref{sec:exp}. Finally, we discuss the consequences of explicit $\cT$ breaking for Fraunhofer patterns and Shapiro steps measurements in twisted junctions in Sec.~\ref{sec:Shapiro} and present concluding remarks in Sec.~\ref{sec:Conclusion}.

\section{Setup and symmetries}
\label{sec:Setup}
We start from the Ginzburg-Landau description of twisted superconductors and derive the form Josephson energy of the twisted interface. While this approach is strictly valid only close to $T_c$, the resulting description of the Josephson energy can be shown to hold also at low temperatures for weak tunneling \cite{Tummuru2021, Volkov2021}.
We can write the Ginzburg-Landau  free energy for twisted bilayer $d$-wave superconductors (omitting gradient terms) as
\begin{align}
    \cF[\psi_1, \psi_2] =& \cF_0[\psi_1] + \cF_0[\psi_2] + A |\psi_1|^2 |\psi_2|^2 \nonumber \\
    &+ B \left( \psi_1^* \psi_2 + h.c. \right) + C \left( \psi_1^{*2} \psi_2^2 + h.c. \right) ,
    \label{eq:free_energy}
\end{align}
where $\psi_{a}$ ($a=1,2$) is the SC order parameter of layer $a$ and $\cF_0[\psi_a] = \alpha |\psi_a|^2 + \beta  |\psi_a|^4$, with $\alpha \sim (T-T_c)$ and $\beta>0$, denotes the free energy of each individual layer. The $B$ term describes single Cooper pair tunnelling between the layers, while $C$ describes Cooper pair co-tunneling. At $\pf$, the former process is forbidden due to a vanishing overlap of the $d$-wave order parameters. Denoting the twist angle as $\theta$, we follow Ref.\ \cite{Can2021} and assume $B=-B_0\cos{2\theta}$ with $B_0>0$  while we take $C$ as constant independent of the twist. 

When $C>0$ the free energy \eqref{eq:free_energy}  admits a $\cT$-broken, chiral SC phase near the $45^{\rm o}$ twist. To see this we assume identical layers and take $\psi_1 = \psi$ and $\psi_2 = \psi e^{i \phi}$ with real $\psi$.  Eq.~\eqref{eq:free_energy} then becomes
\begin{equation}
    \cF(\phi) = \cE_0 - \frac{\hbar}{2e} \left[ J_{c1} \cos 2\theta \cos \phi - \frac{J_{c2}}{2} \cos 2 \phi \right],
    \label{eq:josenergy}
\end{equation}
with $J_{c1}= (4 e/\hbar)  B_0\psi^2$, $J_{c2} =  (8 e / \hbar)  C \psi^4$. Here $\cE_0$ collects terms independent of the phase $\phi$. For sufficiently small $|\cos{2\theta}|$, that is, in the vicinity of $\theta= \pf$, the last term in Eq.\ \eqref{eq:josenergy} begins to dominate and produces a free energy with two non-equivalent minima at $\phi=\pm\phi_0$, resulting in a spontaneously $\cT$-broken superconducting state. The twist angle range for such chiral $\cT$-broken  SC in this model is $\pf \pm \theta_c$ with
\begin{equation}
\theta_c = \frac{1}{2} \arccos \left( \frac{2 J_{c2}}{J_{c1}} \right)   
\label{eq:theta_c}
\end{equation}

Note that $C>0$ is not required by symmetry. Although microscopic calculations often show this to be the case \cite{Tummuru2021,Volkov2021}, including additional effects, such as inhomogeneity, can lead to $C<0$ \cite{Volkov2021}.

Additionally, {we allow for the possibility of \emph{explicit} $\cT$-breaking in the SC state (i.e., we assume that the normal state from which SC emerges may also break $\cT$). Such a possibility is motivated by experimental observations of a memory effect in a few samples, whereby the diode polarity remains robust to thermal and current cycling~\cite{Zhao2023}}. In such a situation, the Ginzburg-Landau free energy will also include a term of the form 
\begin{equation}
\cF_{m}= \frac{im}{2} \left(\psi_1 \psi_2^* - \psi_1^* \psi_2 \right) = m \psi^2\sin{\phi} .
\label{eq:josenergy2}
\end{equation}
Due to the $d$-wave nature of the order parameters $\psi_1$ and $\psi_2$, the factor $m$ in Eq.~\ref{eq:josenergy2} transforms as the irreducible representation $A_2$ of either of the point groups $D_4$ (valid for generic non-zero twist angle) and $D_{4d}$ (valid for $45^{\rm o}$ twist). Additionally, $m$ is odd under time reversal, 
\begin{equation}
\cT: m\to -m.
\label{eq:tm}
\end{equation}
Therefore, $m$ transforms as an out-of-plane magnetization. As this term couples directly to the superconducting phase difference $\phi$ we name it ``magneto-chiral coupling". Note that, unlike the actual out-of-plane magnetic field, this term allows for homogeneous order parameters and does not imply generation of vortices. In principle, magnetic field below $H_{c1}$ would satisfy this requirement, as orbital or spin magnetisation. In this work, we will not consider the influence of Abrikosov vortex physics on the diode effect appropriate for experiments without applied field (see, however, Ref. \cite{Ghosh2022}). An interplay between the magneto-chiral effect identified here and vortex physics would be an interesting topic for future works.

Precisely at $\pf$, the point group $D_{4d}$ does not contain true inversion, but instead a mirror and 8-fold rotation $S_8$ that sends $\psi_1 \rightarrow \psi_2$ and $\psi_2 \rightarrow -\psi_1$, under which $m$ is even. However, at $0^\circ$ or $90^\circ$ twist angle $m$ instead transforms as the irrep $A_{2u}$ of the point group $D_{4h}$, which is odd both under  inversion and a mirror symmetry that interchanges the two layers, $\psi_1 \leftrightarrow \psi_2$. Therefore $m$ cannot arise from an out-of-plane magnetization at $0^\circ$ or $90^\circ$. The simplest (i.e.\ the lowest harmonic) twist-angle dependence consistent with the above requirements is 
\begin{equation}
m = m_0 \sin{2 \theta}, 
\label{eq:mth}
\end{equation}
and we will use this functional form in our considerations below unless otherwise noted. The Josephson current through the junction $I_J(\phi) = (2e/\hbar) \partial_\phi \cF$ is obtained as 
\begin{equation}
I_J(\phi)  = J_{c1}(\theta) \sin{\phi} - J_{c2} \sin{2\phi} + J_m(\theta) \cos{\phi} ,
\label{eq:critcur}
\end{equation}
where we introduced a shorthand notation 
\begin{equation}\label{eq:IJ}
J_{c1}(\theta)= J_{c1}\cos{2\theta},\ \ 
J_m(\theta)= J_m\sin{2\theta},
\end{equation}
and $J_m=(2e/\hbar)\psi^2 m_0$. Interestingly, the same form of the current-phase relation can arise in ferromagnetic Josephson junctions~\cite{Goldobin2007, Goldobin2011, Goldobin2013, Menditto2016, Sickinger2012, goldobin2012_exp}. In this work we will focus on physical aspects that are unique to twisted nodal superconductors, including the twist-angle dependence and how to use the diode response to distinguish between spontaneous and explicit TRSB.

Finally, we define the \emph{thermodynamic} critical current
\begin{equation}
I_c^\pm = \max_\phi [ \pm I_J(\phi) ], 
\end{equation}
which is in general different from the actual measured critical current that may depend on the phase dynamics in the junction, as explained below. We note that thermodynamic non-reciprocity, that is $I_c^+\neq I_c^-$, is only possible when $J_m\neq 0$, in accord with the requirement that time reversal must be broken at the level of the free energy in order to observe the diode effect. On the other hand we will show that ``dynamical" non-reciprocity is possible even when $J_m=0$, provided that the free energy has a double-well structure as described by Eq.\ \eqref{eq:josenergy} when $\theta$ is close enough to $\pf$.

\section{Dynamical diode effect from spontaneous $\cT$ breaking}
\label{sec:dyndiode}

We first discuss the possibility of Josephson diode effect for $J_m=0$. In the symmetry broken state -- $J_{c1}(\theta) <2J_{c2}$ in Eq.\ \eqref{eq:josenergy} --  nonzero $\langle\phi\rangle$ breaks both time-reversal and reflection symmetries, which allows for the diode effect. However, the  current \eqref{eq:critcur} still satisfies $I_J(\phi) = -I_J(-\phi)$, which implies that its maximum and minimum have the same absolute value, i.e. it is the same in both directions.

Nevertheless, as we show below, the actual measured critical current is equal to $\pm \max_\phi |I_J(\phi)|$ only when capacitance of the junction can be ignored (i.e. the junction is overdamped), which typically occurs close to $T_c$. The situation changes when capacitive effects are considered. To study these effects, we use RCSJ model for the Josephson junction dynamics \cite{Tinkham,baronepaterno} in which the phase evolution is governed by 
\begin{equation}
\frac{\hbar C}{2e} \partial_{tt} \phi(t) 
+
\frac{\hbar}{2eR} \partial_{t} \phi 
+J_{c1}(\theta) \sin \phi 
-J_{c2}\sin 2 \phi
  =
J_0,
    \label{eq:RSJ0}
\end{equation}
with $R$ the normal resistance of the junction, $C$ its capacitance and $J_0$ represents the bias current. Voltage across the junctions is equal to $\frac{\hbar}{2e}\partial_t \phi$ per the Josephson relation. It is convenient to normalize all currents by $J_{c2}$ and define timescale $t_0={\hbar}/(2eRJ_{c2})$ to obtain a dimensionless equation 
\begin{equation}
 \beta_c   \partial_{\tau\tau} \varphi 
+
\partial_{\tau} \varphi 
    +
    \bar{J}_{c1} \cos(2\theta) \sin{\phi} 
-\sin{2 \phi}
    =
    \bar{J}_0
    \label{eq:RSJ}
\end{equation}
where $\tau=t/t_0$ is the dimensionless time variable. We defined $\bar{J}_{c1} =J_{c1}/J_{c2}$, $\bar{J}_0=J_0/J_{c2}$ and 
\begin{equation}\label{eq:smc}
\beta_c= {2e R^2 C J_{c2}}/{\hbar}
\end{equation}
denotes the Stewart-McCumber parameter.

Eq.\ \eqref{eq:RSJ} can be interpreted as an equation of motion of a phase ``particle'' (also referred to as the RCSJ particle) with an inertial mass $\propto \beta_c$ in a tilted washboard potential 
\begin{equation}
U(\phi) = -\frac{\hbar}{2e}J_0 \phi +\cF(\phi).
\label{eq:u}
\end{equation}
The parameter $\beta_c$ controls the importance of friction: for $\beta_c\gg1$ friction can be ignored, while for $\beta_c\ll1$ inertia can be neglected and motion is purely viscous. In the latter case, if $U(\phi)$ has any local minima, the motion will terminate there. Therefore, when $\beta_c\ll 1$ the measured critical current is equal to the thermodynamic critical current.

From a practical perspective, $\beta_c$ can be controlled by temperature. It is expected to vanish when $T \rightarrow T_c$, as the critical current $J_{c2}$ goes to zero (see Eq.~\ref{eq:smc}). As temperature is lowered, $\beta_c$ increases due to two effects: the critical current $J_{c2}$ becomes larger, while the resistance of the junction $R$ may additionally increase due to the thermal depletion of quasiparticle population~\cite{Tinkham}.

\subsection{Weakly damped junction: $\beta_c\gg1$}
\label{sec:nodamping}

We now consider the case where capacitive effects are dominant, $\beta_c\gg1$. We first ignore the effects of friction, and 
only take into account an infinitesimal damping to ensure a static steady state when the RCSJ particle is trapped in one of the potential wells. Effects of finite damping will be discussed in the following subsection.
\begin{figure}[t]
		\includegraphics[width=\linewidth]{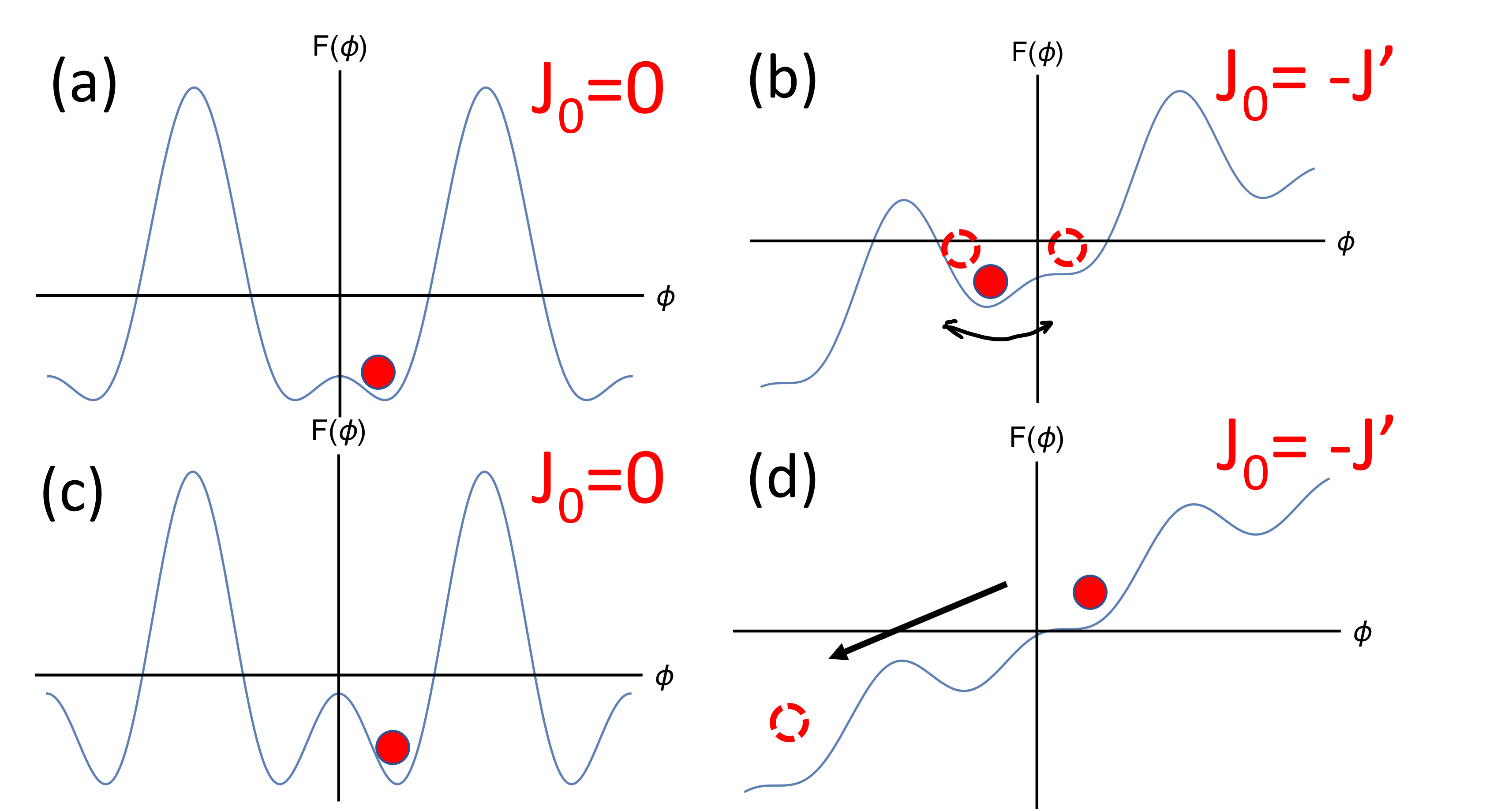}
	\caption{Schematic of the phase dynamics for $-J_0<J_c$ and $\beta_c\gg1$. (a,b) for $2J_{c2}>J_{c1} \cos(2\theta)>0.79J_{c2}$ (see Eq. \eqref{eq:jc1crdiode}); (c,d) for $0.79J_{c2}>J_{c1} \cos(2\theta)$. In (a,b) one observes that $\phi$ can relocate to the other local potential minimum at $J'$, when the original minimum becomes unstable. However, energy conservation allows unbounded motion only in case (d).
	}
	\label{fig:jprime}
\end{figure}

In the absence of friction, the motion can be analyzed from the energy viewpoint. As discussed above, for $J_{c1}(\theta)<2J_{c2}$ the free energy at $J_0=0$ has two distinct minima. This is the interesting case for the onset of JDE. We assume without the loss of generality that the minimum spontaneously chosen is $\phi=\phi_0>0$ as illustrated in Fig.\ \ref{fig:jprime}(a,b). Turning on the bias current corresponds to tilting the potential landscape for the phase particle. 
Because the potential is not symmetric around the chosen minimum, the sign of $J_0$ will matter for the determination of the critical current. It therefore follows that $I_c$ can be different for the two directions of the bias current, giving rise to the junction non-reciprocity.

Let us consider the onset of voltage across the junction upon increasing the bias current $J_0$. In our analysis we assume that the current is turned on adiabatically, i.e.\ at each value of $J_0+dJ_0$ we perform the stability analysis starting from the energy  minimum at the previous step $\phi_0(J_0)$. Clearly, the phase value will track the local minimum of $U(\phi)$, until $J_0=J_c$ where the minimum becomes unstable.

For $J_0>0$ the particle follows the initial minimum at $\phi_0$ until $J_0=J_c$, where it becomes unstable. $J_c$ is given by
\begin{equation}
\begin{gathered}
    J_c = \frac{J_{c2}}{8} \sqrt{8-2 x^2+2x\sqrt{x^2+8}}
    \left(\sqrt{x^2+8}+x\right).
    \end{gathered}
\end{equation}
with $x={J_{c1}(\theta)}/(2J_{c2})$.
For $J_0>J_c$ the phase $\phi$ will exhibit unbounded motion resulting in a nonzero $\dot\phi$ and, therefore, voltage.

The situation is different for $J_0<0$. In this case, the minimum that is adiabatically connected to the original one can become unstable at a lower value of $|J_0| = J' < J_c$ (see Fig. \ref{fig:jprime}). 

\begin{figure}[t]
		\includegraphics[width=\linewidth]{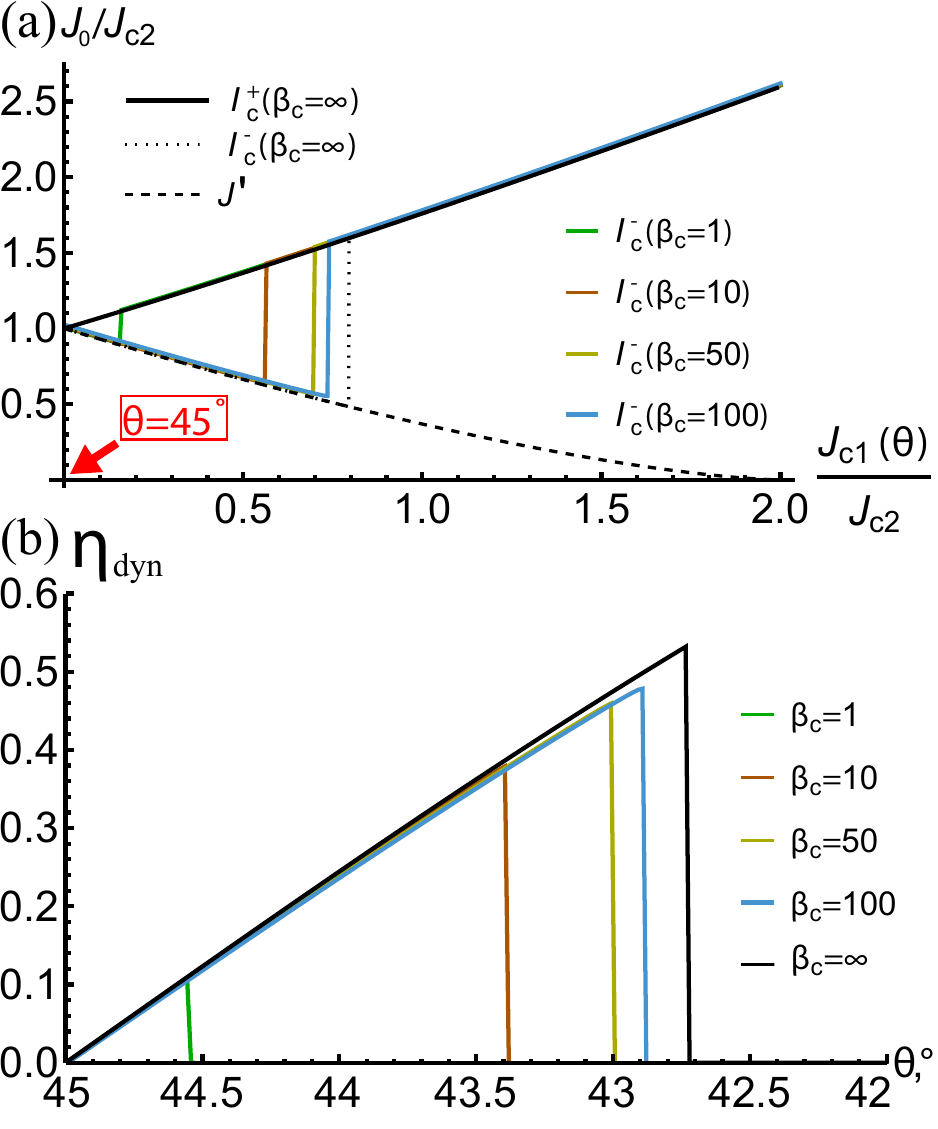}
	\caption{(a) Characteristic currents for the dynamical diode effect from the RCSJ model \eqref{eq:RSJ} as a function of $J_{c1}(\theta)$. $J_{c1}(\theta)=0$ corresponds to $\theta=45^\circ$. Dashed black line shows the  $J_0'$ value, where the initial potential minimum (taken to be at $\phi=\phi_0>0$) becomes unstable (see Fig.\ \ref{fig:jprime}(b)). The black dotted and solid lines show the normalized critical current for positive and negative current bias (see Fig.\ \ref{fig:jprime}(b,c)) in the $\beta_c\to \infty$ limit. Diode effect ($I_c^+\neq I_c^-$) is present for $J_{c1}$ less than a critical value \eqref{eq:jc1crdiode}. Colored lines represent $I_c^-$ from the numerical solution of Eq.\ \eqref{eq:RSJ} for finite $\beta_c$ ($I_c^+$ remains same as for $\beta_c\to\infty$). The transition from diode to non-diode regime remains abrupt for all $\beta_c$, but shifts to lower $J_{c1}(\theta)$ values. (b) Diode efficiency $\eta_{dyn}$, Eq. \eqref{eq:etadyn} as a function of twist angle $\theta$ for $J_{c1}(\theta=0)=0.1 J_{c2}$. Note that the system is in TRSB state already for $\theta>39.2^\circ$ for these values. 
}
	\label{fig:diode2nd}
\end{figure}

The value of $J'$ can be found as follows. Since the minimum becomes unstable at this current value, the second derivative of $U(\phi)$, Eq. \eqref{eq:u},  should be zero at the respective equilibrium position, $U''(\phi')=0$. Noting that $U''(\phi)$ is independent of $J_0$ we can find $\phi'$ that corresponds to the second local minimum disappearing. The corresponding current value ($J'$) for its disappearance is then given by
\begin{equation}
    J'=\sqrt{1-\cos^2(\phi'_-)}(J_{c1}(\theta)-2J_{c2} \cos(\phi'_-)).
\end{equation}
The dependence of $J'$ on $J_{c1}(\theta)/J_{c2}$ is shown in Fig.\ \ref{fig:diode2nd} together with that of $J_c$. One observes that for $J_{c1}(\theta)=2J_{c2}$ one has $J'=0$, whereas for  $J_{c1}(\theta)=0$, one obtains $J'=J_c$.

For $|J|>|J'|$ two cases can be distinguished. Fig. \ref{fig:jprime}(b) illustrates the case when at $J'$ the initial energy is not sufficient for $\phi$ to overcome the next potential hump. Therefore, $\phi$ will exhibit an oscillatory motion around the remaining minimum. Infinitesimal damping will eventually localize the particle at the remaining stable minimum. So the actual critical current in this case remains $J_c$ for both directions of applied current and the behavior is reciprocal.

The other possibility is shown in \ref{fig:jprime}(d). In this case, the potential energy of $\phi$ is sufficient to overcome the next barrier. This results, for infinitesimal damping, in an unbounded motion and non-zero voltage. In this case the critical current is $-J'$ rather than $-J_c$ and therefore, there is a nonzero diode effect for a given initial $\phi_0$ value and adiabatic current ramping. Analyzing Eq.\ \eqref{eq:u} numerically, we find that this case is realized for 
\begin{equation}
    \left(\frac{J_{c1}(\theta)}{J_{c2}}\right)<\left(\frac{J_{c1}(\theta)}{J_{c2}}\right)_{D} \approx 0.79.
    \label{eq:jc1crdiode}
\end{equation}
Thus, importantly, we conclude that the diode effect that occurs in the absence of explicit time-reversal breaking is only expected within a portion of the TRSB phase. In particular, this sets a limit for the twist angle $|\theta - 45^\circ|\lesssim 0.4|\theta_{\rm TRSB} - 45^\circ|$ for the onset of the diode effect. In addition, JDE must vanish at  $\theta=\pf$ despite the broken time reversal symmetry at $\pf$. We therefore stress that its observation is a sufficient but not necessary condition for TRSB. For $0.79J_{c2}<J_{c1}(\theta)<2J_{c2}$, TRS is broken but the diode effect cannot be observed due to kinetic energy arguments.

\subsection{Intermediate $\beta_c$}

Above we demonstrated that for $\beta_c\to\infty$ in Eq. \eqref{eq:RSJ0}, there is a finite diode effect for small enough $J_{c1}(\theta)$. At the same time, for $\beta_c=0$ diode effect should be absent. Here we study numerically the behavior of Eq.\ \eqref{eq:RSJ0} for finite values of $\beta_c$.

To find the critical current, we assume the initial conditions to be $\phi(0)=\phi_0$; $\phi'(0) = 0$ and solve  Eq.\ \eqref{eq:RSJ} with bias current ramping up linearly in time, $\bar{J}_0 = \tau/\tau_0$, where $\tau_0$ is the inverse ramping rate. The critical current is determined by the value of $\tau$ when $|\phi|$ becomes larger than $\pi$. This criterion implies that $\phi$ has traversed the largest potential barrier (see Fig.\ \ref{fig:jprime}) and has enough energy to continue moving. The criterion yields the exact value of $J_c$ only in the limit $\tau_0\to \infty$  but we have checked that already for $\tau_0=4000$ the result is independent of $\tau_0$ and use these converged results in our discussion.

In Fig. \ref{fig:diode2nd} (a) we present the value of the critical current for reverse bias (as shown in Fig. \ref{fig:jprime}) as a function of the second harmonic critical current for different values of $\beta_c$. We can also quantify the dynamical diode efficiency by
\begin{equation}
    \eta_{\rm dyn} = \frac{I_M^+-I_M^-}{I_M^++I_M^-}
    \label{eq:etadyn}
\end{equation}
where $M=L(R)$ for $J_m>(<)0$. In Fig. \ref{fig:diode2nd} (b) we show the calculated $\eta_{dyn}$ as a function of twist angle. As expected from the above discussion, it vanishes at $45^\circ$ exactly, then increases up to a finite value on decreasing the twist angle, vanishing at a critical value. Note that this critical value always remains within the TRSB phase.

\begin{figure}
	\includegraphics[width=\linewidth]{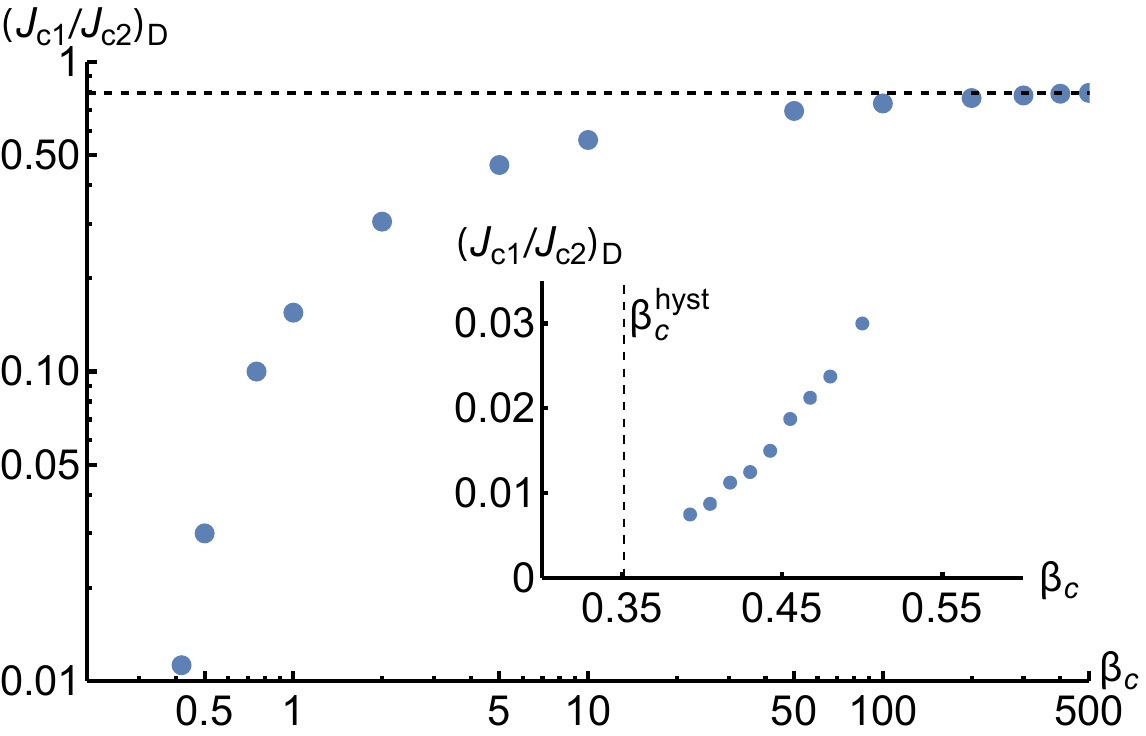}
	\caption{Critical value of the first harmonic critical current for the onset of the diode effect. For large McCumber parameter $\beta_c$, the result agrees with the analytical $\beta_c\to \infty$ limit Eq. \eqref{eq:jc1crdiode}. Inset shows the behavior near the onset of hysteretic behavior of the RCSJ model $\beta_c^{\rm hyst}$.
 }
	\label{fig:diode2ndcritcur}
\end{figure}

Let us now discuss the effects of decreasing $\beta_c$ (increasing damping)
One observes that the transition between the diode and reciprocal regime remains abrupt down to the lowest value considered. 
This allows us to define a critical value of the first Josephson harmonic $J_{c1}(\theta)$ for the onset of the diode effect, as shown in Fig.\  \ref{fig:diode2ndcritcur}. From the inset, one observes that the diode effect becomes extremely fragile at small values of $\beta_c$.

At low $\beta_c$ the RCSJ model solutions are known to become non-hysteretic \cite{mccumber1968effect}. In the hysteretic regime, the switching current $I_c$ (critical current on increasing the current from zero) and retrapping current $I_r$ (where the voltage drops to zero on decreasing the current from high value) are different, $I_r<I_c$. For the current bias between $I_c$ and $I_r$, therefore, two steady state solutions of Eq.\ \eqref{eq:RSJ} are possible. In the non-hysteretic regime, $I_c=I_r$, and there is only one steady state solution of Eq.\ \eqref{eq:RSJ} for each value of $J_0$. However, the dynamical diode effect requires that there can be a zero-voltage and a finite-voltage solution at the same value of the bias current, depending on the initial condition for $\phi$. Thus, hysteresis is a necessary condition for the dynamical diode effect.

Fig. \ref{fig:diode2ndcritcur}, inset, suggests that the diode effect disappears at around $\beta_c \approx 0.35-0.4$. This critical value coincides with the hysteresis onset of the RCSJ model with the second harmonic current phase relation for $J_{c1}(\theta)=0$. This value can be obtained from the value $\beta_c^{\rm hyst}\approx 0.7$ \cite{mccumber1968effect,zorin2005josephson} for the first harmonic RCSJ model by replacing $2\phi \to \tilde{\phi}$ in Eq.~\eqref{eq:RSJ}.

\section{Diode effects in the presence of explicit $\cT$ breaking}
\label{sec:trsb}

We now consider the situation where time-reversal symmetry is explicitly broken -- that is, the GL free energy includes the contribution $\cF_m$ defined in Eq.\ \eqref{eq:josenergy2}, with non-vanishing $\cT$-breaking field $m$. The $\cF_m$ term is such that $I_J(\phi) \neq - I_J(-\phi)$, and therefore leads to a Josephson diode effect already at the level of thermodynamic critical currents. Such a $\cT$-breaking term, if it survives for $T>T_c$, could also give rise to memory effects when the system is heated up above $T_c$. The junction would then show the same polarity of the diode effect after cycling above the critical temperature or after being driven into the resistive state by exceeding its critical current.

\subsection{Thermodynamic diode effect}
\label{sec:thermdiode}
\begin{figure*}[t]
	\includegraphics[width=\textwidth]{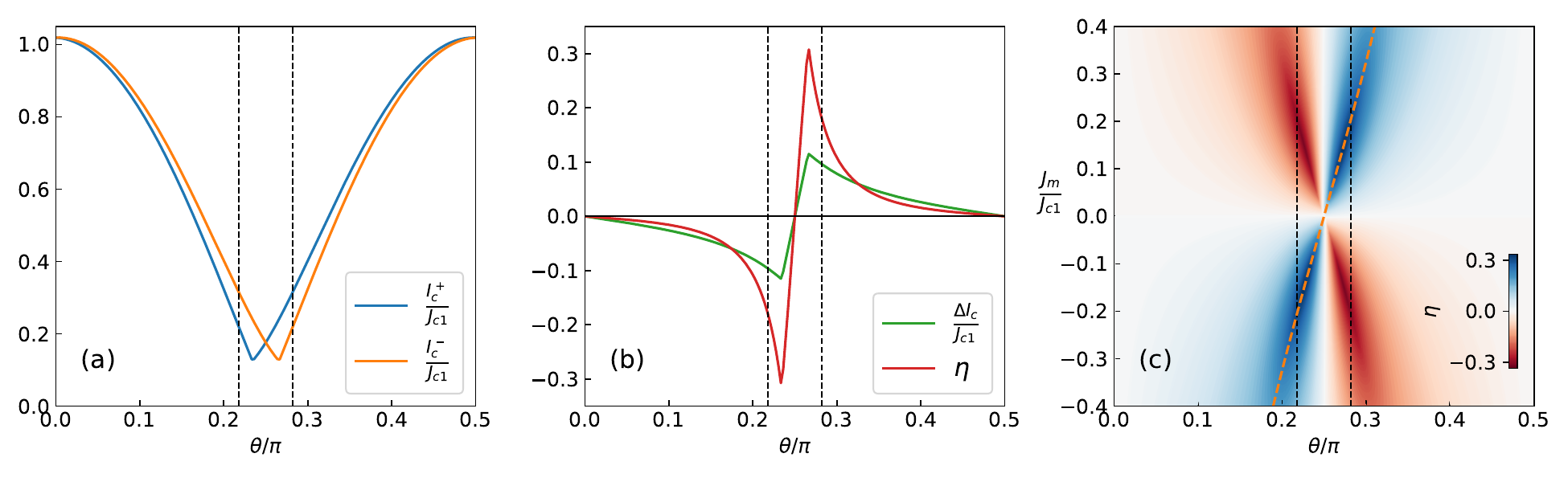}
	\caption{Thermodynamic Josephson diode effect as a function of twist angle $\theta$, obtained from the Ginzburg-Landau description in Eq.~\ref{eq:critcur} (equivalent to the over-damped regime $\beta_c \ll 1$ of the RCSJ model). We take $J_{c2} = 0.1 J_{c1}$, which sets the critical twist angle for the appearance of topological superconductivity to $\theta_c = 39.2^\circ$ through Eq.~\ref{eq:theta_c}, denoted by vertical black dashed lines. (a): Critical Josephson currents $I_c^\pm$ develop an asymmetry in the presence of the $\cT$-breaking term $J_m$ (here $J_m = \sqrt{2} J_{c2}$). (b): The corresponding difference $\Delta I_c=I_c^+ - I_c^-$ and figure of merit $\eta$, defined in Eq.\ \eqref{eq:eff}, exhibit a sharp increase upon entering the spontaneous $\cal T$-breaking phase. The diode effect vanishes at $\theta=45^\circ$, where the diode polarity reverses. (c) The figure of merit $\eta$ in the $\theta-J_m$ plane. The maximal diode efficiency $| \eta | = 1/3$ lies on the $ J_m / J_{c1} = \pm \cot 2\theta$  curves (the orange dashed line shows the ``$-$" solution).}
	\label{fig2}
\end{figure*}

 For concreteness we assume the simplest twist angle dependence compatible with the symmetry constraints of the magneto-chiral coupling indicated in Eq.\ \eqref{eq:mth}.  However, our results are not changed qualitatively by adding higher harmonics that are also allowed by symmetry. It is then straightforward to obtain the thermodynamic critical currents, defined as $I_c^\pm = {\rm max}_\phi \left[ \pm I_J(\phi) \right]$, from the GL free energy description that leads to Eq.\ \eqref{eq:critcur} for the Josephson current. Fig.~\ref{fig2} shows that when both $J_{c2}$ and $J_m$ are non-zero the junction exhibits non-reciprocal behavior, $I_c^+\neq I_c^-$, for all twist angles except for $\theta=0,45^\circ$.

 This result can be understood as follows. First, note that for any $m \neq 0$ the free energy component $\cF_m$ is odd under both $\cT$ and $\cP$, and hence the basic symmetry requirements for the SC diode effect are met. As already discussed in Sec.\ II, for an untwisted junction symmetry dictates that when $m=0$, the free energy is symmetric around $\phi=0$ and there is no thermodynamic non-reciprocity. At $\theta=45^\circ$ the magneto-chiral coupling is maximal, as per Eq.\ \eqref{eq:mth}, but now the first-order Josephson term $J_{c1}(\theta)$ vanishes. It is easy to verify that $\cF(\phi)$ is then symmetric about its $\cT$-breaking minima at $\phi_0=\pm \pi/2$ and, once again, this implies equal thermodynamic critical currents for both directions. 

Furthermore, when $J_{c2}=0$ (no Cooper pair co-tunneling), $I_J$ is stationary for $ \phi_0 = \arctan \left[J_{c1}(\theta)/J_{m}(\theta) \right]$ and $I_c^\pm = \max \left[ \pm I_J(\phi_0) \right]$ with
\begin{equation}
    I_J(\phi_0) = \pm \sqrt{ J^2_{c1}(\theta) + J_m^2(\theta)} ,
\end{equation}
and there is again no critical current asymmetry. This occurs because $I_J(\phi)$, while no longer anti-symmetric in $\phi$, is nevertheless anti-symmetric with respect to a shifted origin. We thus need all three of $J_{c1}(\theta)$, $J_{c2}$ and $J_m(\theta)$ non-vanishing to observe thermodynamic non-reciprocity, as also demonstrated numerically in Fig.~\ref{fig2}.

We can derive a bound on the diode efficiency, expressed through the figure of merit  
\begin{equation}\label{eq:eff}
\eta = \frac{I_c^+ - I_c^-}{I_c^+ + I_c^-} ,
\end{equation} 
as follows. We first observe that the maximum of $ | \eta |$ occurs along the $J_m(\theta) = \pm J_{c1}(\theta)$ lines, Fig.\ \ref{fig2} (c).  This can be understood by recasting Eq.\ \eqref{eq:critcur} using trigonometric identities as
\begin{equation}
I_J(\phi)  = \sqrt{ J^2_{c1}(\theta) + J_m^2(\theta)} \sin(\phi + \alpha ) - J_{c2} \sin{2 \phi},
\label{eq:critcur2}
\end{equation}
where $\tan{\alpha}=J_m(\theta)/J_{c1}(\theta)$. Clearly, the current $I_c^+$ will be maximal when the maxima of the two sine functions coincide. This happens when $\alpha = - \pi/4$ or $\alpha=3 \pi/4$, implying $J_m(\theta) = - J_{c1}(\theta)$. By sketching the two sine functions it is also easy to see that, at the same time, such $\alpha$ choices minimize $I_c^-$, hence leading to maximum $\eta$ for $J_m(\theta) = - J_{c1}(\theta)$, as indeed found numerically~\footnote{Similarly, when $\alpha=+\pi/4$ or $\alpha=-3\pi/4$ the minima of the two sine functions in Eq.~\ref{eq:critcur2} are aligned, leading to optimal but reversed diode efficiency, $\eta = -1/3$. Changing the sign of $J_{c2}$ also has the effect of reversing the diode efficiency.}. 

Taking $\alpha=- \pi/4$  the current-phase relation \eqref{eq:critcur2} simplifies to $I_J(\phi) = J_m(\theta) (\cos \phi - \sin \phi) - J_{c2} \sin 2 \phi$ which gives the following critical currents
\begin{equation}
    I_c^+ = \sqrt{2} J_m(\theta) + J_{c2},
\end{equation} 
and
\begin{equation}
    I_c^- = 
    \begin{cases}
    \sqrt{2} J_m(\theta) - J_{c2}, & J_{c2} < \frac{J_m(\theta)}{2\sqrt{2}}, \\
    \frac{J_m^2(\theta)}{4 J_{c2}} + J_{c2}, & J_{c2} > \frac{J_m(\theta)}{2\sqrt{2}}.
    \label{eq:Ic_minus_m}
    \end{cases}
\end{equation}
The maximal diode efficiency $\eta = 1/3$ occurs when $J_{c2} = J_m(\theta)/\sqrt{2}$, as shown in Fig.~\ref{fig2} (c).
Note that at $J_{c2} = \sqrt{2} |J_m(\theta)|$ the free energy transitions from having a double-well structure (for large $J_{c2}$) to a $\cT$-broken single-well structure (for small $J_{c2}$) -- see also Fig.~\ref{fig:trsb_jc1cr}. The largest diode efficiency therefore occurs in the regime with a single $\cT$-broken ground state. In the double-well regime the largest diode efficiency is $\eta = 7/25$ near the transition point $J_{c2} = \sqrt{2} J_m(\theta)$.

\begin{figure*}[t]
	\includegraphics[width=\textwidth]{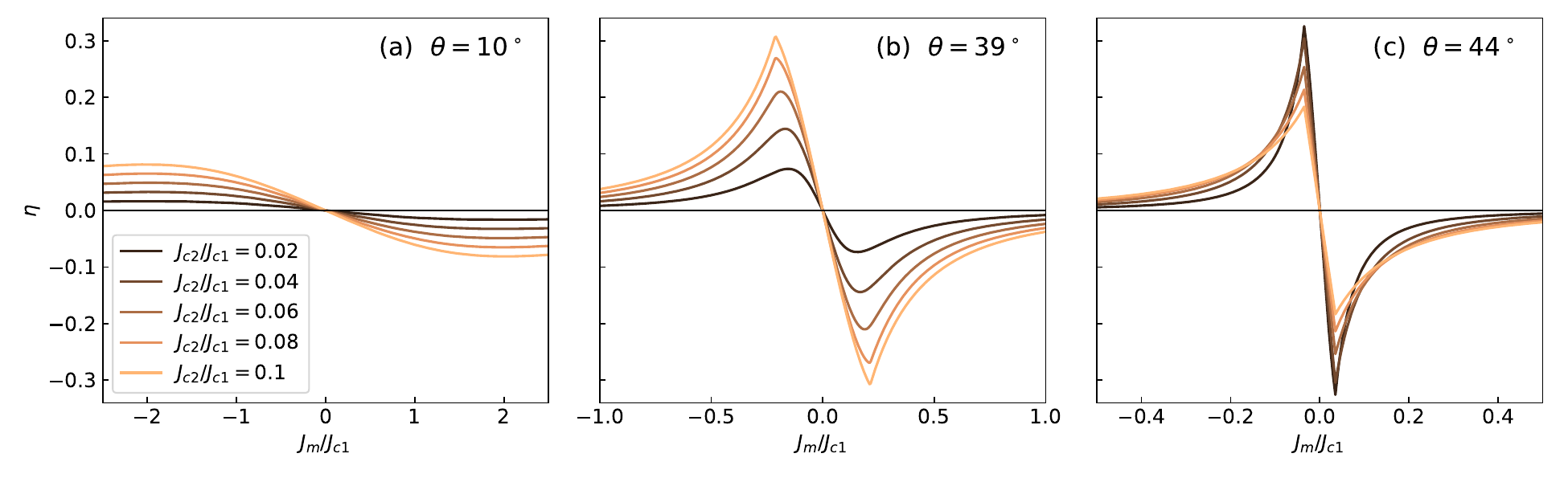}
	\caption{Thermodynamic Josephson diode effect as a function of the $\cT$-breaking perturbation $J_m$, obtained from the Ginzburg-Landau description in Eq.~\ref{eq:critcur} (equivalent to the over-damped regime $\beta_c \ll 1$ of the RCSJ model). Various curves denote $J_{c2}/J_{c1}$ between $0$ and $0.1$, corresponding to $\theta_c$ ranging from $45^{\rm o}$ (vanishing topological region) to $39.2^\circ$. (a) For small twist angles, the efficiency parameter $\eta$ shows a broad and low peak centered around a large optimal value $J^{\rm opt}_m$. (b) and (c): For larger twist angles near $\pf$, $\eta$ peaks near its theoretical maximum of $1/3$ for the value of $J_{c2}/J_{c1}$ that best approximates $\theta_c \approx \theta$ through Eq.~\ref{eq:theta_c}. The $\eta$ peaks become sharper, and occur at a decreasing optimal value $J_m^{\rm opt}$, when $\theta$ approaches $\pf$.}
	\label{fig3}
\end{figure*}

Fig.~\ref{fig3} further illustrates the non-monotonic behavior of the asymmetry parameter $\eta$ as function of $J_m$, with a peak at an optimal value $J_m^{\rm opt}$. This peak is very broad for low twist-angle junctions, and becomes progressively sharper when $\theta$ approaches $45^\circ$. 
Junctions close to $45^\circ$ twist, as in Fig.~\ref{fig3} (c), are very sensitive to explicit $\cT$-breaking, as shown by their small optimal value $J_m^{\rm opt}$.

We also stress that, within the GL free energy description, both the theory with $J_{c2} > 0$ (where the system exhibits chiral SC in a range of twist angles) and $J_{c2} < 0$ (which favors the trivial phase for all twist angles) show similar phenomenology for the thermodynamic diode effect. As such, the observation of a non-zero asymmetry $\eta$ in the presence of explicitly-broken $\cT$ is, in itself, insufficient to determine whether the underlying superconductor is chiral. However, if the $\cT$-breaking term can be controlled externally (such as by applying an out-of-plane magnetic field), the presence of two $\cT$-breaking ground states for $J_{c2} > 0$ should be accompanied by a hysteresis loop when $J_m$  is swept back and forth around $0$. Such a hysteresis is not expected in the theory with $J_{c2}<0$, which supports a single $\cT$-preserving ground state.

\subsection{Dynamical diode effect in the presence of explicit time-reversal breaking}

\begin{figure}
\includegraphics[width=\linewidth]{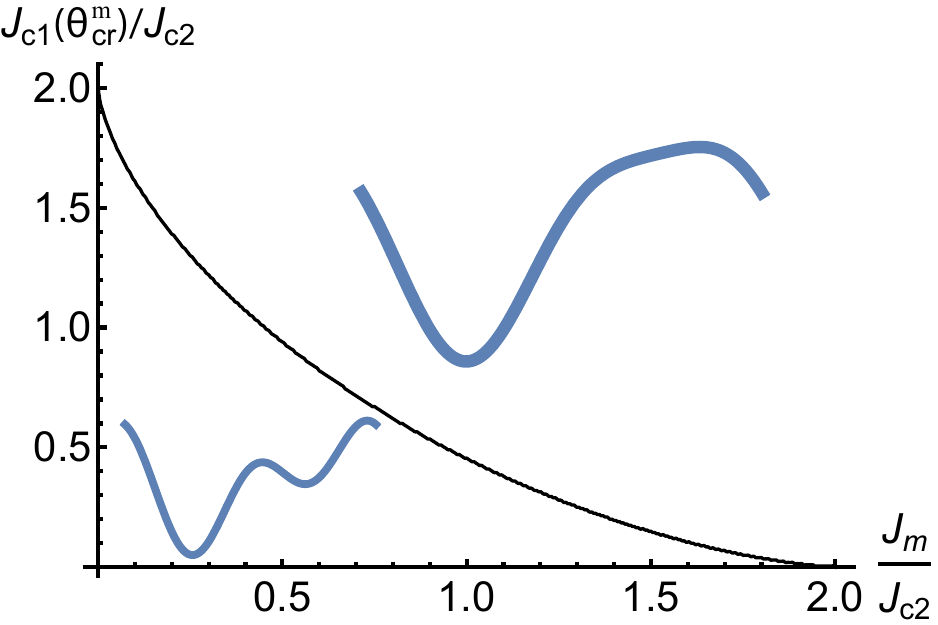}
	\caption{Critical value of the first harmonic Josephson current $J_{c1}(\theta_{cr}^m)$ for the emergence of two potential minima in the Josephson energy $U_m(\phi)$ \eqref{eq:um} for $J_0=0$ (illustrated by insets) as a function of TRSB Josephson current $J_m$.}
\label{fig:trsb_jc1cr}
\end{figure}

We now revisit the results of Sec. \ref{sec:dyndiode} in the presence of explicit TRSB terms in the Josephson energy, which now takes the form
\begin{equation}
U_m(\phi) = -\frac{\hbar}{2e}J_0 \phi 
+\frac{\hbar}{2e} J_m(\theta) \sin \phi
+\cF(\phi),
\label{eq:um}
\end{equation}
where $\cF(\phi)$ is defined by Eq.\ \eqref{eq:josenergy}.
For $J_0=0$, there is always only one global minimum of Eq. \eqref{eq:um}. However, two distinct local minima in the Josephson energy $U_m(\phi)$ can still remain. In particular, for a fixed value of $J_m$, two minima exist for $J_{c1}(\theta)$ ranging from zero (corresponding to $45^\circ$) to a finite critical value.
In Fig. \ref{fig:trsb_jc1cr} we present this critical value of the first harmonic of the Josephson current  as a function of $J_m$. Note that for $J_m>2J_{c2}$ only a single minimum exists for all values of $J_{c1}$.
\\

\begin{figure}
\includegraphics[width=\linewidth]{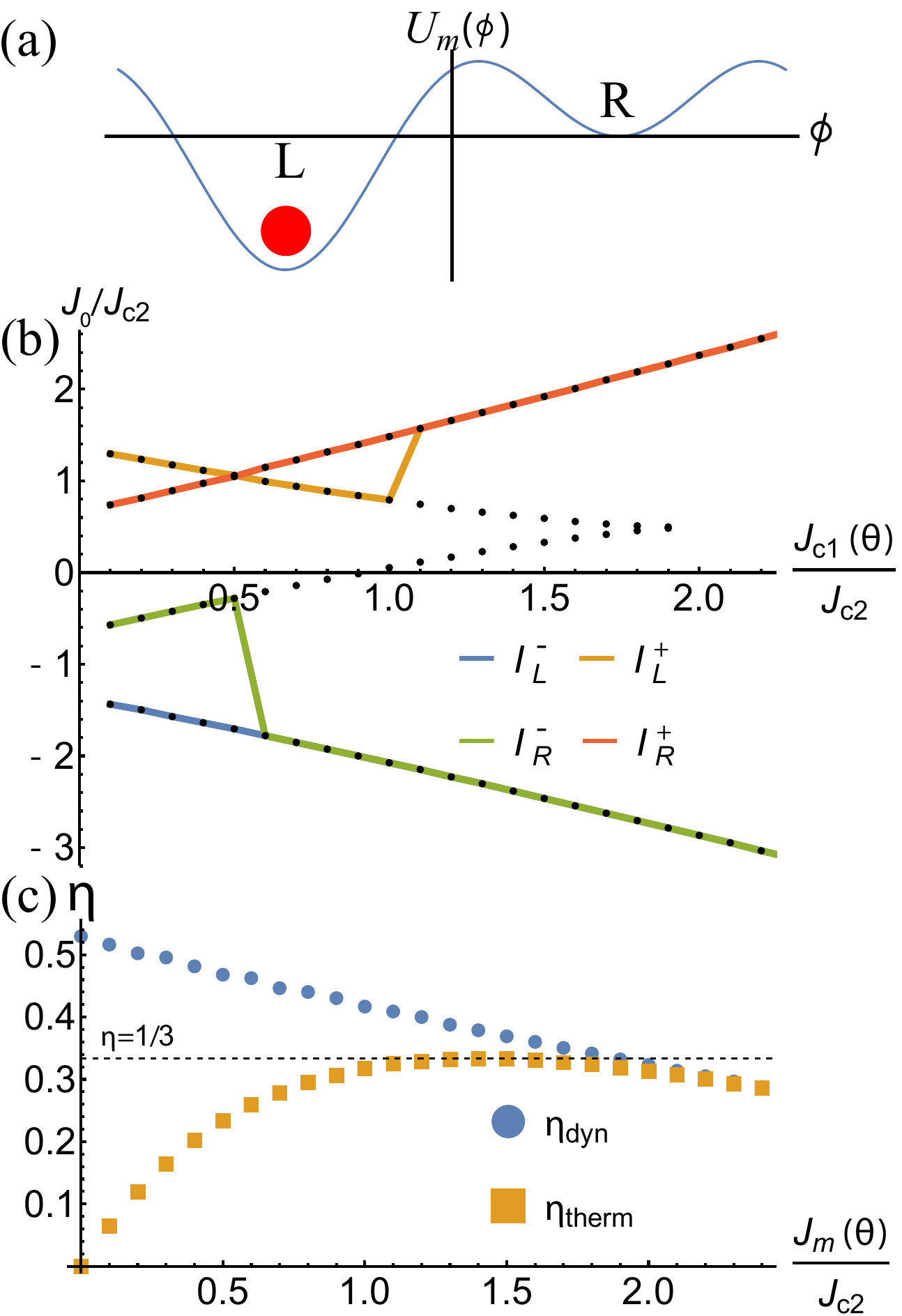}
	\caption{(a) Illustration of the Josephson potential \eqref{eq:um} with TRSB term $J_m<J_{c2}$ in the absence of external current. 
 (b) Four characteristic current values $I_c^{L,R, \pm}$ for $\beta_c\to \infty$ and $J_m/J_{c2} = 0.5$. The colored lines show the values where the $\phi$ can actually escape the local minimum, leading to the onset of the voltage. The black points mark all values where one of the local minima ceases to exist, not necessarily leading to the voltage onset.
 (c) Maximum diode efficiency (for varying $J_{c1}(\theta)$) for the thermodynamic and dynamic diode effect.
 }
\label{fig:trsb_char_cur}
\end{figure}

In this section, similarly to Sec. \ref{sec:nodamping}, we consider the dynamical diode effect neglecting damping, i.e, for $\beta_c\gg1$ and assuming adiabatic current ramping. Let us consider the case when two distinct minima exist (Fig.\ \ref{fig:trsb_char_cur}(a)). For $\phi$ being initially at L (R) there exist two characteristic current values ($I_{L(R)}^\pm$) when the particle escapes that minimum under positive/negative current bias. In Fig. \ref{fig:trsb_char_cur} (b) we present $I_{L(R)}^\pm$ calculated numerically for $J_m/J_{c2}=0.5$ as a function of $J_{c1}$. The behavior qualitatively resembles Fig.\ \ref{fig:diode2nd}, but the curves for $I_{L}$ and $I_R$ are not symmetric, reflecting the explicit TRSB. 

The black dots in Fig. \ref{fig:trsb_char_cur}(b) mark the values where one of the minima ceases to exist, but not necessarily leading to the onset of voltage. At low $J_{c1}$ values there are four such characteristic currents, two positive and two negative. For the current values between the second and third point, two minima in the potential exist. Remarkably, for $J_{c1}(\theta)\geq J_{c2}$ one observes that this range does not cover zero current. This implies that additional minima in \eqref{eq:um} can be generated by the applied current in presence of finite $J_m$, even if there is only one minimum at $J_0=0$ (see Fig. \ref{fig:trsb_jc1cr}). The application of a current can thus restore, to a degree, the symmetry of the potential, by counteracting the TRSB term in Eq. \eqref{eq:um}.

Let us now discuss the expected behavior of the critical current in an experiment. In thermodynamic equilibrium, $\phi$ always starts at $L$ for $J_m>0$ (Fig. \ref{fig:trsb_char_cur} (a)), and therefore $I_R$ and bistability of the potential is unobservable in the critical current (see, however, Sec.\ \ref{sec:exp} for the non-equilibrium case). Nonetheless, the possibility of preemptive escape alters the diode effect strength with respect to thermodynamic effect corresponding to the overdamped limit discussed in Sec.\ \ref{sec:thermdiode}.

In Fig.\ \ref{fig:trsb_char_cur}(c) we present the maximal diode efficiency $\eta_{\rm dyn}$, Eq. \eqref{eq:etadyn}, for varying $J_{c1}$ and fixed $J_m$ and compare it with the thermodynamic $\eta_{}$, Eq.\ \eqref{eq:eff}, where $|I_c^{\pm}| = \max[|I_L^\pm|,|I_R^\pm|] $.
The dynamical diode efficiency can be larger than the maximal value $1/3$ for $\eta_{}$, due to the possibility of early escape. At low values of $J_m$, $\eta_{\rm dyn}\approx0.53$, corresponding to the dynamical diode effect caused by spontaneous symmetry breaking, Fig. \ref{fig:diode2nd}. 

\section{Dynamical vs. thermodynamic diode effect: experimental protocol}
\label{sec:exp}

In the above sections we have demonstrated that both spontaneous (Sec. \ref{sec:dyndiode}) and explicit (Sec. \ref{sec:trsb}) TRSB can lead to current nonreciprocity in twist junctions of $d$-wave superconductors. However, for purely spontaneous TRSB, one expects a randomly fluctuating sign of the nonreciprocity, while the addition of even a small explicit TRSB term will fix it. This raises the question of how the possible spontaneous nature of TRSB and the bistability of the current-phase relation can be identified in an experiment. As has been discussed above, while the diode effect in the thermodynamic critical current requires the second harmonic of the Josephson current to be present, it may occur even in the case when there is no bistability.
Here we explain how can one characterize the TRSB in twisted nodal superconductors by adapting the protocol used in experiments on ferromagnetic Josephson  
junctions \cite{goldobin2012_exp}. 

\subsection{Spontaneous TRSB: $J_m=0$}

Since TRSB occurs spontaneously, the equilibrium value $\phi$ is chosen randomly to be equal to $+\phi_0$ or $-\phi_0$. The key insight is that one can deterministically prepare the system in one or the other equilibrium by current sweeping.

Consider adiabatically decreasing $|J_0|$ from high bias larger than $I_c$ towards zero. For $\beta_c>\beta_c^{cr}$ (hysteretic regime) the voltage will only go to zero at $I_r$, $|I_r|<I_c$. Let us focus on how retrapping occurs. The value of $I_r$ corresponds to the case when $\phi(t)$ eventually stops ($\dot\phi=0$ as $t\to\infty$) for any initial conditions. For such a solution to exist, the potential, Eq. \eqref{eq:u} has to have at least one local minimum (otherwise, there will be a force acting on $\phi$ and causing motion). In Sec. \ref{sec:dyndiode}, we demonstrated (see Fig. \ref{fig:jprime}) that for $J'<J_0<J_c$ values only one minimum exists. Moreover, this minimum is adiabatically connected with the right (left) minimum at $J_0=0$ for $J_0>(<)0$. Thus, for $I_r>J'$ (which requires sufficiently strong damping, but still in the hysteretic regime), the value of $\phi$ can be deterministically prepared by retrapping.

A more general argument can be given, that extends to lower $I_r$, where for each interval $\phi\in [2\pi n, 2\pi (n+1)]$ there exist two minima of \eqref{eq:u}. Without loss of generality, let us assume $J_0>0$. For $J_0>I_r$, $\dot\phi(t)$ is a periodic function such that $\phi(t)$ advances by $2\pi$ over the period. This implies that $\dot\phi$ has a global minimum $\dot\phi_{\rm min}$ for every period. For large $J_0$, where Josephson nonlinearity can be neglected $\dot\phi>0$ for all $t$. Therefore, for large enough $J_0$, $\dot\phi_{\rm min}>0$. To go into the retrapped state, $\dot\phi_{\rm min}$ has to go through zero at some value of $J_0$. We will show now that this value is the retrapping current. At this critical value, $\dot\phi_{\rm min}=0$, but $\dot\phi\geq0$ at all times. The equation \eqref{eq:u}, on the other hand, implies that $\ddot\phi\sim -U'(\phi)$. So if $\dot\phi(t_{\rm min})=0$, $\dot\phi$ has to be negative either before or after this moment unless $U_m'(\phi(t_{\rm min}))=0$. This implies that $\phi$ has to be in the minimum or maximum of $U_m(\phi)$. The former can be excluded, since infinitesimal increase of $J_0$ will not set $\phi$ into motion. Thus, $\phi$ is at the maximum of $U_m(\phi)$. This implies that reducing $J_0$ further would lead to $\dot\phi<0$. Since energy is dissipated with time, $\phi$ will never be able to overcome the potential maximum and is thus retrapped. For Eq. \eqref{eq:u}, only one such maximum exists per $2\pi$ interval if $J_{c1}\neq0$. Therefore, $\phi$ would stop to the left (right) of the potential maximum for $J_0>(<)0$ and will be trapped in different minima depending on the sign of $J_0$. Note that for large $\beta_c$, $\phi$ can perform many oscillations before stopping, and thus the result of retrapping becomes extremely sensitive to $\beta_c$ \cite{Goldobin2013}.

\begin{figure}[h!]
\includegraphics[width=\linewidth]{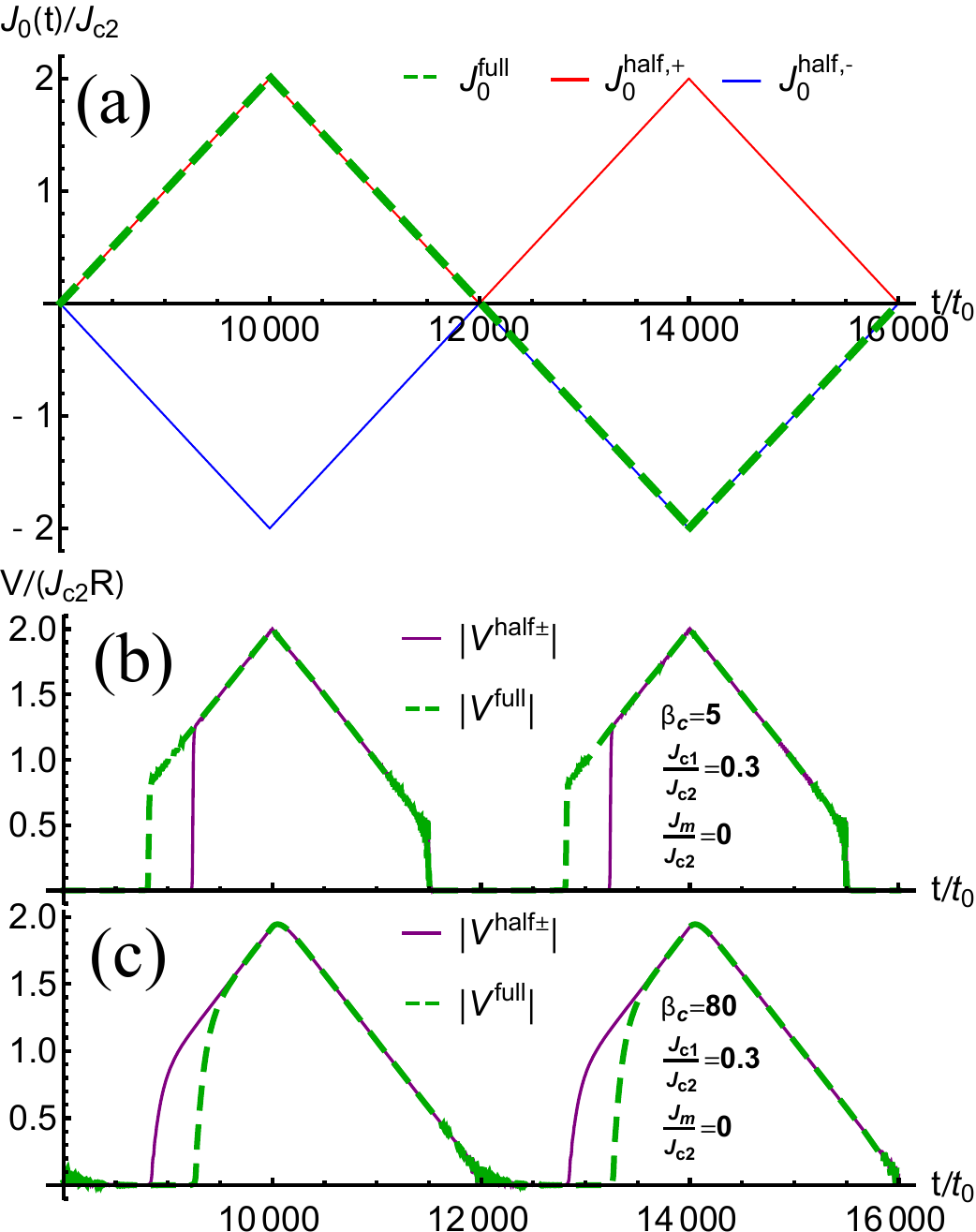}
	\caption{(a) Three current sweeping protocols, Eq. \eqref{eq:prot} for $t_c=2000 t_0$. (b,c) Voltage (averaged over $10 t_0$) from numerical solution of Eq. \eqref{eq:RSJ} for $\beta_c = 5$ (b) and $\beta_c = 80$ (c).  Voltage is shown for the second cycle of the current sweeping to suppress the effects of initial conditions. The time at which voltage first appears corresponds to a critical current value in panel (a). The two half-sweep directions (purple line) lead to identical critical current amplitudes, as is expected in the absence of explicit TRSB. For the full sweep protocol (green dashed line), on the other hand, the current is sweeped in the opposite direction with respect to the position of the phase at zero current (see Fig.  \ref{fig:jprime}). The difference in full sweep and half sweep current demonstrates the bistability of the Josephson energy.
	}
	\label{fig:tdep1}
\end{figure}

The above discussion leads to the following experimental protocol. One can ``prepare" $\phi$ to be in the left or right minimum by retrapping. Then, biasing the junction with positive or negative voltage will lead to different critical currents in a part of TRSB phase (see Fig. \ref{fig:diode2ndcritcur}). One can implement these ideas by comparing the measured voltage for three periodic current patterns depicted in Fig.\ \ref{fig:tdep1}(a). We refer to these as full sweep  and half sweep protocols $J_0^{\rm full}(t)$ and $J_0^{{\rm half},\pm}(t)$, mathematically defined as 
\begin{equation}
\begin{gathered}
        J_0^{\rm full}(t) = 
    \begin{cases}
        J_{\rm max}s(t), &0<s(t)<1\\
       J_{\rm max}(2-s(t))  &1<s(t)<3\\
       J_{\rm max}(-4+s(t))  &3<s(t)<4
    \end{cases}
\end{gathered}
\label{eq:prot}
\end{equation}
where $s(t)=\mod[t,4t_c]/t_c$ and $J_0^{{\rm half},\pm}(t)=\pm \left | J_0^{\rm full}(t)\right |$.
For the full sweep and half sweep the junction is biased in opposite directions after an interval over which they coincide. The latter can be viewed as a "preparation" step, where $\phi$ is prepared in the same minimum, but afterwards biased in opposite directions for full and half sweep. A difference in the measured voltage between the two protocols implies dynamical diode effect and spontaneous TRSB.

To demonstrate the protocol explicitly we numerically solved Eq. \eqref{eq:RSJ}. Two typical solutions are presented in Fig. \ref{fig:tdep1} (b,c). One observes that full and half sweep protocols always give different critical current values, as expected from dynamical diode effect. Note that there is no difference between $\pm$ half sweep cases, because there is no explicit TRSB. One observes that changing the value of $\beta_c$ leads to different relation between full and half sweep cases. In particular, this points out that the minimum, where $\phi$ is retrapped depends on the value of $\beta_c$ \cite{Goldobin2013}. Nonetheless, for both cases the dynamical diode effect and thus spontaneous TRSB can be demonstrated from this protocol.
\begin{figure}[h!]
\includegraphics[width=\linewidth]{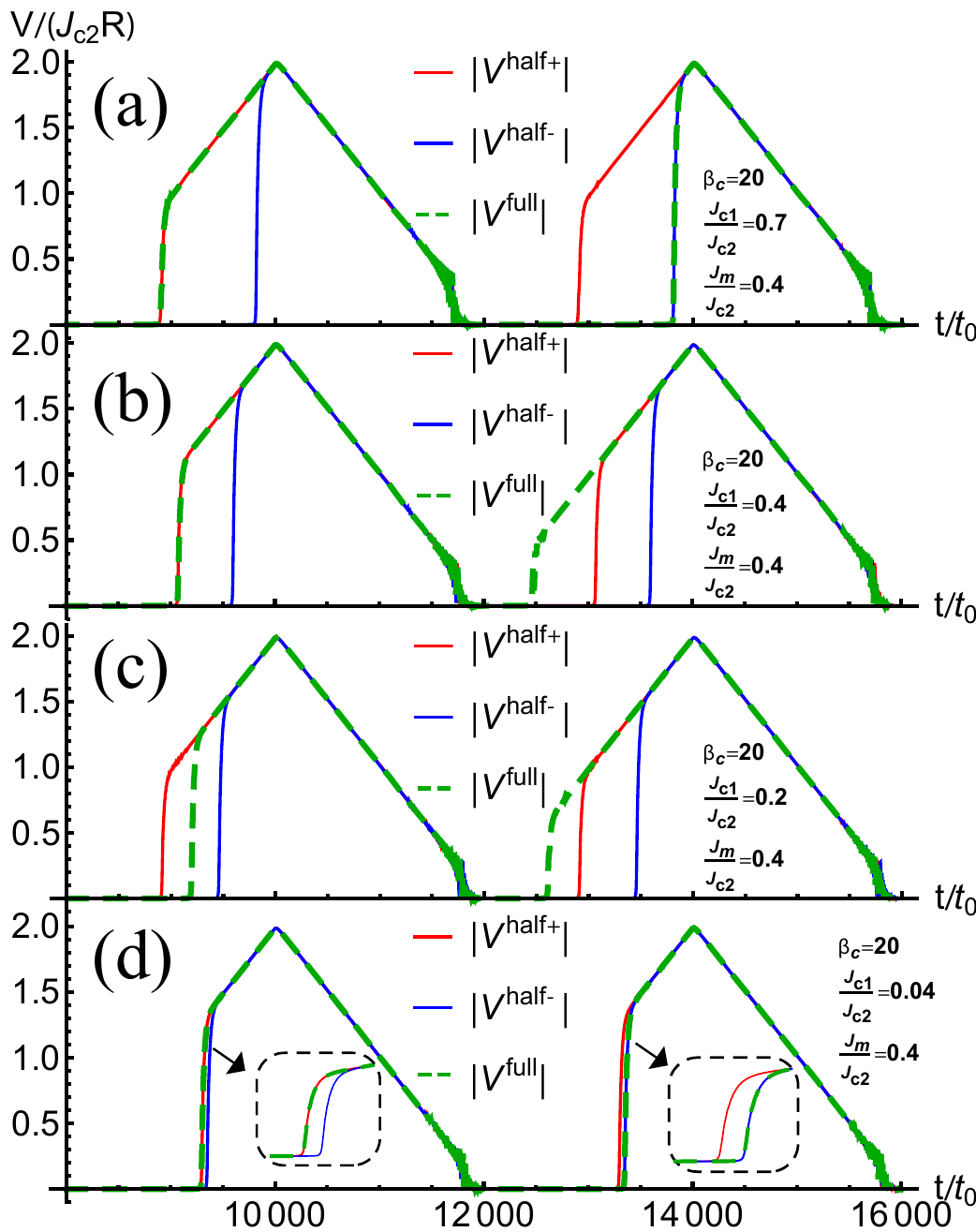}
	\caption{(a-d) Voltage (averaged over $10 t_0$) from numerical solution of Eq. \eqref{eq:RSJ} with added TRSB term $J_m$ (values given in the panels) and current driving shown in Fig. \ref{fig:tdep1} (a). Insets in (d) marked by black dashed line show magnified transition region (not to scale). In contrast to Fig. \ref{fig:tdep1}, the half-sweep critical current (related to the voltage onset time) depends on the current polarity (red solid line for "+", blue solid line for "-"), reflecting the explicit TRSB. The full-sweep critical current (green dashed line) coincides with one of the two half-sweep values when the first harmonic is large (a) or extremely small (d). For other values of $J_{c1}$ (b,c), the full sweep shows values of critical current distinct from both (c) or at least one (b) of the half-sweep protocols. The latter observation can be used to demonstrate bistability of the Josephson potential even if TRSB is explicitly broken.
	}
	\label{fig:tdep2}
\end{figure}

\subsection{Explicit TRSB: $J_m\neq0$}
\label{sec:exp:trsb}

 Large $J_m$  excludes the possibility of deterministic retrapping since there is only one minimum at zero current bias, see Fig.\ \ref{fig:trsb_jc1cr}). However, at smaller values of $J_m$, two minima remnant from the spontaneous TRSB state can still exist.

In Fig. \ref{fig:tdep2} we present four characteristic regimes that occur as a function of $J_{c1}$ for a fixed $J_m$. For large $J_{c1}$ (Fig.\ \ref{fig:tdep2}(a)), the full and half sweep protocols yield the same values of the critical current. However, positive and negative bias (two stages of full sweep, or $\pm$ half-sweep protocols) show different critical current for voltage onset. This is the thermodynamic diode effect, described in Sec.\ \ref{sec:thermdiode}. Decreasing $J_{c1}$ (Fig.\ \ref{fig:tdep2}(b,c)) leads to the splitting between full and half sweep protocols, consistent with additional characteristic current values appearing in Fig.\ \ref{fig:trsb_char_cur}(b). 
 This indicates bistability of the potential \eqref{eq:um} remnant from the spontaneous TRSB state at $J_m=0$. Interestingly, the full-half splitting disappears at yet smaller values of $J_{c1}$ (corresponding to twist angles closest to $45^\circ$), Fig.\ \ref{fig:tdep2}(d). The analysis of the numerical solution suggest that at low $J_{c1}$ retrapping training is ineffective - $\phi$ always gets trapped in the global minimum, leaving only signatures of the thermodynamic diode effect.

Thus, the described protocol gives access (for sufficiently large $\beta_c$ and not too close to $45^\circ$) to all four characteristic values of the current (Fig. \ref{fig:tdep2} (b,c)). This allows to directly demonstrate the presence of two minima in the Josephson energy even in the presence of explicit TRSB.


\begin{figure*}
    \centering
	\includegraphics[width=\textwidth]{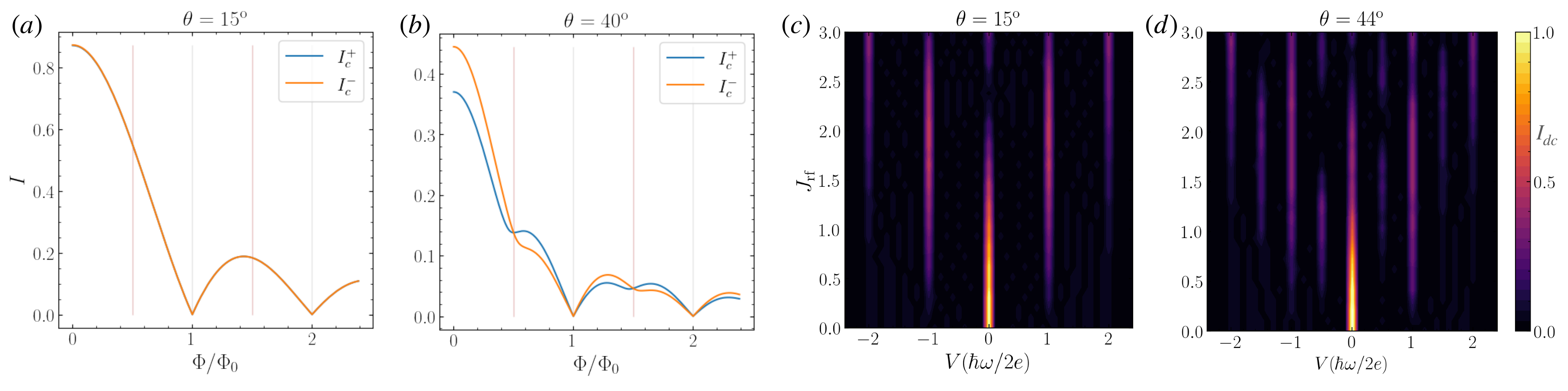}
	\caption{Fraunhofer interference patterns and Shapiro steps and the JDE. (a,b) In the presence of a dominant second harmonic and $J_m= 0.25 J_{c2}$ (with $J_{c2}=0.2$ in units of $J_{c1}$), the envelopes of the positive and negative critical currents are different. At half-integer flux quanta, the curves intersect and the diode effect vanishes because the contribution from the second harmonic goes to zero due to its $\pi$ periodicity. (c,d) Close to $\pf$, the normalized Shapiro step heights (indicated with the color scale) are asymmetric about zero voltage.}
	\label{fig7}
\end{figure*}

\section{Other implications of second harmonic and magneto-chiral coupling}
\label{sec:Shapiro}

As discussed in earlier theoretical works \cite{Tummuru2021,Volkov2021}, and confirmed by experimental observations in twisted BSCCO bilayers \cite{Zhao2021}, the presence of the second harmonic term $J_{c2}(\theta)$ in the free energy $\cF(\phi)$ can be probed by means of perturbing the junction using in-plane magnetic field or a radio-frequency (RF) drive. In this Section we briefly discuss the effect of these perturbations on the Josephson response in the presence of both $J_{c2}(\theta)$ and magneto-chiral coupling $J_{m}(\theta)$. We find that the asymmetry between the two current directions is sensitive to the presence of in-plane magnetic field which can therefore be used to probe the effect in greater detail. The magneto-chiral coupling, on the other hand, modifies the junction response to the RF drive and produces asymmetry in the resulting Shapiro steps.

\subsection{Fraunhofer patterns}

In a regular Josephson junction, all of the junction area can support the critical current $I_c^{\pm}$ because of the position independent phase difference between the superconductors. An in-plane magnetic field, however, induces a phase gradient such that the maximum and minimum interlayer current densities vary spatially. More concretely, with the junction plane perpendicular to the $z$ direction, when a magnetic field $B_y$ is applied along $y$, the Josephson current density along $x$ given by $I_J(\phi_x)$, where the phase variation \cite{Tinkham, baronepaterno}
\begin{equation}
    \phi_x = \frac{2 \pi d}{\Phi_0} B_y x + \phi_0.
\end{equation}
Here $d$ is the effective junction thickness, $\Phi_0= (hc/2e)$ denotes the superconducting flux quantum and $\phi_0$ is a uniform phase shift. When integrated over the area of a junction of unit length $L$ and width $W$, the interference from different contributions results in a Fraunhofer pattern
\begin{equation}
\label{eq:fraun}
\begin{gathered}
    I_J(\phi_0, \Phi) =
     \left[J_{c1}(\theta) \sin(\phi_0)+J_m(\theta)\cos(\phi_0)\right] \frac{\sin{( \pi\Phi}/{\Phi_0})}{\pi\Phi/{\Phi_0}} \\
     - J_{c2} \sin(2\phi_0) \frac{\sin{(2 \pi\Phi}/{\Phi_0})}{{2 \pi\Phi}/{\Phi_0}},
\end{gathered}
\end{equation}
where $\Phi=dLB_y$ is the magnetic flux through the junction. 
The equilibrium critical currents for a given flux are determined by the extrema of $I_J(\phi_0, \Phi)$ with respect to $\phi_0$.

We notice that Eq.\ \eqref{eq:fraun} has the same form as the zero-field expression \cref{eq:critcur}, albeit with renormalized coefficients. The second and first harmonics are renormalized differently. In particular, at ${\Phi}/{\Phi_0} = n+1/2$ with $n$ integer, the second harmonic vanishes, while the first harmonic and the magneto-chiral term do not. At the corresponding field values $B_y$ we therefore expect the thermodynamic diode effect to be suppressed. This is indeed observed in our simulation results \cref{fig7}(a,b).
The diode effect persists at nonzero field strengths except at half-integer fluxes where contribution from the second harmonic vanishes because of its $\pi$ periodicity. In its absence, the diode effect cannot be induced by the $J_m$ term alone and, hence, $I_c^{+} = I_c^{-}$. Moreover, the switch in the values of the second harmonic at half-integer fluxes changes the diode polarity and is manifested as the oscillating pattern in the $I_c^{+}$ and $I_c^{-}$ curves. The polarity flipping as a function of the in-plane field is, therefore, suggestive of a dominant second harmonic alongside the magneto-chiral term.

\subsection{Shapiro steps}

When a Josephson junction is subjected to an external RF drive, the $I$-$V$ curves show steps at integer multiples of the voltage $V_s = (\hbar\omega / 2e)$, where $\omega$ is the drive frequency. The phenomenon is captured by the resistively shunted junction (RSJ) model \cite{baronepaterno}
\begin{equation}
    \frac{\hbar}{2 e R} \frac{\partial \phi}{\partial t} + I_{J}(\phi) = I_{\rm{dc}} + J_{\rm{rf}} \sin(\omega t),
    \label{eq:rsj}
\end{equation}
where $R$ is the junction resistance, $I_{\rm{dc}}$ is the measured direct current (dc) and $J_{\rm{rf}}$ is the drive amplitude. We solve Eq.\ \eqref{eq:rsj} numerically using energy units where $\hbar/2e = 1$, $R=0.7$ and $\omega=0.6$. Two representative results are depicted in \cref{fig7}(c-d). The $n$-th Shapiro step represents a $n$ photon-assisted tunneling of Cooper pairs across the junction. In the presence of a dominant second harmonic, co-tunneling of Cooper pairs gives rise to steps at half-integer multiples of $V_s$. In a symmetric Josephson junction, the step heights for positive and negative bias voltages are identical. We observe that in the presence of the magneto-chiral term  the step heights at half-integer voltages are no longer symmetric about zero voltage, \cref{fig7}(d). Such an asymmetry, therefore, could be indicative of explicit TRSB in the junction.


\section{Discussion and conclusions}
\label{sec:Conclusion}

The free energy $\cF(\phi)$ of a Josephson junction can develop the characteristic double-well structure when the Cooper pair co-tunneling process (i.e.\ the second CPR harmonic $J_{c2}$) becomes dominant. In twisted $c$-axis junctions between two $d$-wave superconductors this is expected to occur as the twist angle approaches $45^\circ$ and the single-pair tunneling $J_{c1}$ is suppressed \cite{Can2021}. In this work we identified two types of Josephson diode effects that can occur in these kinds of junctions.

The {\em dynamical} JDE depends on the initial state of the system and relies on the fact that, for a given free energy minimum, the barrier between the SC and resistive states is generally different for the two polarities of the bias current. Therefore, when the damping is small, the junction can exhibit consistent JDE, provided that the measurement protocol is devised such that one always starts from the same free energy minimum. According to our analysis, spontaneous TRSB is a prerequisite for the dynamical JDE; however, it occurs only in a portion of the $\cT$-broken phase {as illustrated in Fig.~\ref{fig:setup}}. Observation of the dynamical JDE therefore provides evidence for the bistability of the free energy landscape and spontaneous $\cT$-breaking in the twisted junction.

The {\em thermodynamic} JDE by contrast depends on explicit $\cT$ breaking that is present at the level of the GL free energy -- that is, exists already in the normal state of the material. In this case $\cF(\phi)$ has a single global minimum away from $\phi=0,\pi$ and the diode effect exhibits a fixed polarity controlled by the sign of the $\cT$ breaking term. As we discussed, the second harmonic $J_{c2}$ must be present in this case also for the device to show JDE, even though it need not be dominant. In addition, thermodynamic JDE survives in the limit of strong damping. Importantly, signatures of the bistability of the potential can be revealed in an experiment even in the case when $\cT$ is broken explicitly. As discussed in Sec.\ \ref{sec:exp:trsb} this can be achieved by comparing different current sweep protocols, allowing to trap the phase in a local minimum of the free energy.        

Some consequences of the spontaneous $\cT$-breaking predicted by theory \cite{Can2021,Tummuru2021,Volkov2021} have been explored in earlier work by Zhao {\em et al.}~\cite{Zhao2021} who reported anomalous Fraunhofer diffraction patterns and half-integer Shapiro steps in near-$45^\circ$ twisted BSCCO junctions. More recently, the same group reported evidence of {\em zero-field} SC diode effect in these devices \cite{Zhao2023}. Samples with twist angle slightly away from $45^\circ$ (but within a window of about $\pm 6^\circ$) showed behavior consistent with the dynamical diode effect discussed in Sec.\ \ref{sec:dyndiode}, indicative of a Josephson free energy with a pronounced double-minimum structure. The non-reciprocal response was probed via the ``full-sweep/half-sweep protocol" described in Sec.\ \ref{sec:exp}, whereby the current sequence applied to the twisted junction is defined so as to controllably prepare the system in one of the two $\cT$-broken minima. Importantly, junctions outside of the $45\pm 6^\circ$ twist angle window showed reciprocal behavior, consistent with our result that dynamical JDE is only expected to be observed within a portion of the TRSB phase. Furthermore, the diode effect was found to vanish in the limit $\theta\to 45^\circ$, in accord with the discussion in Sec.\ III. Our theory therefore provides a good explanatory framework for the experimental findings of Zhao {\em et al.}~\cite{Zhao2021,Zhao2023}. Together, these works make a compelling case for spontaneous TRSB in high-quality BSCCO junctions with a twist angle close to $45^\circ$. 

Several samples studied in Ref.\ \cite{Zhao2021} were reported to exhibit a memory effect indicative of the thermodynamic JDE, with fixed diode polarity independent of its current bias history \cite{Zhao2023}. Most of them showed the full sweep/half sweep splitting, indicating bistability of the potential, see Sec. \ref{sec:exp:trsb}. According to our analysis in Sec.\ \ref{sec:trsb} such behavior is suggestive of explicit $\cT$-breaking in the device that is, presumably, present already in the normal state. Since optimally doped BSCCO crystals are normally thought to be non-magnetic, the nature of this normal-state $\cT$-breaking poses an interesting open question. We see two distinct possibilities for its origin: (i) Even though a single monolayer BSCCO is non-magnetic, it is possible that a twisted bilayer develops normal state orbital or spin magnetism. This would not be unprecedented -- twisted graphene bilayers are well known to develop $\cT$-breaking instabilities even thought a single monolayer is non-magnetic. (ii) A `vestigial' order~\cite{Fernandes2019} that arises from fluctuations of the two superconducting order parameters breaking time reversal symmetry \cite{babaev2013,grinenko2021state,Liu2023, maccari2023}. Briefly, the idea is that above $T_c$ the individual phase-averaged order parameters vanish, $\langle\psi_1\rangle=\langle\psi_2\rangle=0$, but the composite object $m\propto \langle i\psi_1\psi_2^*+{\rm c.c.}\rangle$ may remain ordered up to a higher critical temperature $T_m>T_c$. This scenario requires the relative phase $\phi$ to remain fixed at one or the other $\cT$-breaking value even above $T_c$, hence enabling the return to the same free energy minimum upon cooling back below $T_c$. We note however, that in this scenario, below $T_c$ current training of the diode polarity should be possible, as no additional degrees of freedom are generated. Therefore, coupling of superconducting fluctuations to the other degrees of freedom (spins or orbitals) is required to explain the memory effect.

Josephson diode effects have also been observed in twisted BSCCO junctions by two other groups~\cite{Ghosh2022, Zhu2023}. Ref.~\cite{Ghosh2022} reports a diode effect in presence of a magnetic field along $z$. Its efficiency was found to be largest near $45^{\rm o}$, but nevertheless non-zero for all twist angles considered. Further, the diode polarity can be switched by cycling an applied out-of-plane magnetic field, accompanied by a hysteresis loop. Such observations have been attributed to the presence of a component of the magnetic field through the junction -- generated by the in-plane bending of flux lines connecting the misaligned Abrikosov vortex lattices of the two BSCCO flakes. Therefore, in such a setup the identification of potential signatures of spontaneous $\cT$-breaking, as discussed in the present work, will be much more subtle. Developing a generalization of our theory to include effects of vortices remains an interesting question for future work. Ref.~\cite{Zhu2023}, by contrast, reports observation of the diode effect in the nominal absence of external magnetic field for a sample close to $45^{\rm o}$. We note that our theory predicts that the thermodynamic diode efficiency rises rapidly as the angle is tuned away from 45$^\circ$ (see Fig. \ref{fig2}) and even a small misalignment from $45^{\rm o}$ can lead to a significant diode effect. The finite value of $I_c$ in the sample has been interpreted in Ref.~\cite{Zhu2023} as evidence of non-$d$-wave pairing component. However, the remnant critical current could as well arise from the second harmonic mechanism considered in our work \cite{Tummuru2021, Volkov2021}. Shapiro step measurements could be potentially used to test this scenario.

 Our results are agnostic to the microscopic origin of the second-harmonic term in the Josephson free energy. A first possibility is that of direct Cooper pair co-tunnelling between the twisted BSSCO flakes \cite{Can2021}. The expected magnitude of such a contribution is however difficult to compute accurately \cite{Tummuru2021,Haenel2022}. Another interesting recent proposal is that twist-angle inhomogeneities for near--$45^{\circ}$ junctions may lead to a first harmonic term that vanishes on average but still retains significant spatial fluctuations~\cite{Yuan2023inhomogeneityinduced}. Such a setup was then shown to lead, under suitable conditions, to an effective second-harmonic term with the correct sign to promote spontaneous $\cT$-breaking at the interface. Similarly, we do not attempt to identify the microscopic mechanism behind the possible residual $\cT$-breaking suggested in the normal state of twisted BSCCO bilayers by the results of Ref.\ \cite{Zhao2023}. This is clearly an intriguing effect and, provided that it can be reproduced in more samples and that conditions for its onset are better understood, furnishes an interesting topic for future studies. 
 
 On the other hand, the phenomenological character of our model allows the application of our results to other systems where the same current-phase relation appears. For example, our results for the undamped and overdamped limit are in agreement with Refs. \cite{xiao2023probing} and \cite{hu2023}, respectively. Therefore, the analysis performed in this work for arbitrary damping and the experimental protocols discussed in Sec. \ref{sec:exp} can prove useful for the analysis of a wide variety of systems \cite{Goldobin2007,hu2023,xiao2023probing}.

\section*{Acknowledgments}

We thank X.~Cui, P.~Kim and S.~Y.~F.~Zhao for insightful discussions. \'E. L.-H. acknowledges
support from the Gordon and Betty Moore Foundation’s EPiQS Initiative, Grant
GBMF8682 at Caltech. T.~T. acknowledges funding from the European Research Council (ERC) under the European Union’s Horizon 2020 research and innovation programm (ERC-StG-Neupert-757867-PARATOP).
J.H.P.\ is partially supported by the Air Force Office of Scientific Research under Grant No.~FA9550-20-1-0136, the NSF CAREER Grant No. DMR-1941569, and the Alfred P. Sloan Foundation through a Sloan Research Fellowship. The Flatiron Institute is a
division of the Simons Foundation. Work at UBC (M.F.) was supported by Natural Sciences and Engineering Research Council of Canada (NSERC),  Canada First Research Excellence Fund (CFREF), and Canadian Institute for Advanced Research (CIFAR). Portion of the work has been completed at The Aspen Center for Physics. 

\bibliography{ref}

\end{document}